\documentclass[pra,reprint,10pt]{revtex4}
\usepackage{graphicx}

\usepackage{times}
\usepackage{amsmath}
\usepackage{amssymb}
\usepackage{float}

\begin{document}

\title{Coherent backscattering of light from ultracold
and optically dense atomic ensembles}

\author{I.M. Sokolov${}^{1,2}$ and D.V. Kupriyanov${}^{1}$}

\affiliation{$^{1}$Department of Theoretical Physics, State Polytechnic
University, 195251, St.-Petersburg, Russia \\ \small $^{2}$Institute for
Analytical Instrumentation, Russian Academy of Sciences, 198103, St.-Petersburg, Russia  }

\author{C.I. Sukenik and M.D. Havey}

\affiliation{Old Dominion University, Department of Physics, Norfolk, Virginia 23529 \
}

\begin{abstract} We review experimental and theoretical studies of
coherent backscattering of near resonant radiation from an
ultracold atomic gas in the weak localization regime. Recent
accomplishments in high resolution spectroscopy of atomic
ensembles based on the coherent backscattering process are
discussed. We also propose several new experimental schemes for
time-dependent spectroscopy as applied to multiple scattering in
the regime of weak localization.
\end{abstract}



\maketitle%


\section{Introduction}
The study of optical phenomena is an ancient quantitative
scientific discipline, with historical roots extending more than
two millennia into the past.  It was therefore remarkable that, as
recently as twenty years ago, in 1984, a completely new classical
optics effect was reported for the first time in the scientific
literature.  It was then that Ishimaru and Kuga \cite{Ishimaru}
reported the observation of coherent backscattering (CBS) of light
from a disordered scattering sample.  This report was quickly
followed by experimental and theoretical work \cite{Wolf,Albada}
that included explanation of the effect based on the classical
physical optics of electromagnetic wave scattering in a disordered
medium. In the coherent backscattering effect, there is an
enhancement in the intensity of light scattered in the nearly
backwards direction from a liquid or optically disordered solid.
The enhancement may be as large as a factor of two over the
usually expected average incoherent scattered light intensity.
Further, the enhancement effect is concentrated in a cone-shaped
interference profile typically a few milliradians in angular
width. The fundamental mechanism developed was that
electromagnetic wave scattering along reciprocal, or
time-reversed, multiple scattering paths preserves the relative
phase. The result of phase preservation, after configuration
averaging, along the reciprocal paths leads directly to prediction
of the main features of the coherent backscattering cone for semi
infinite disordered scattering samples.

The basic coherent backscattering effect is remarkably robust, and
can be observed in wave scattering from a wide range of common
natural and manmade materials \cite{Sheng,POAN,Mish}. In the
optical regime, the quantitative features are also quite sensitive
to the polarization of the incident and detected waves.  The CBS
effect is not restricted to scattering of electromagnetic waves,
but has been observed, for example, in acoustics, ultrasonics, and
in propagation of waves in the solid earth.  For each of these
areas there has been a significant range of fundamental studies
and development of practical applications \cite{Sheng,POAN},
particularly in the areas of imaging or detection of embedded
objects within diffusive media. Another very important associated
research area is connected to lasing and wave amplification in
random media having gain \cite{HuiCao1}.

It is our purpose in this review to consider the more specialized
and more recent developments associated with coherent
backscattering of light in ultracold atomic gaseous samples. Such
samples have characteristically very high Q and strong optical
resonances, making them unique in comparison to scattering from
classical condensed or liquid samples \cite{LagTig}.  Further, for
scattering from single atoms, the varied influences of optical and
atomic polarization, and the responses of atomic systems to
applied static and dynamic fields are well known in the linear
response regime \cite{Louden}. Nonlinear responses to applied
electromagnetic fields are also well studied
\cite{ChTnDpRocGr,Mollow}. These characteristics make theoretical
and experimental study of mesoscopic wave scattering in atomic
media an attractive and accessible area of research.

Mesoscale processes in dense and cold atomic vapors can also
display a constellation of fundamental and potentially important
phenomena. One of these, strong localization of light
\cite{Sheng}, is a dynamic research area in its own right.  Strong
localization of light is the optical analog of Anderson
localization of electrons \cite{Anderson}, in which energy
transport through a medium is suppressed by interferences mediated
by multiple scattering in a spatially random medium.  Two reports
of strong localization in condensed systems have been made in the
literature, one in the optical regime \cite{Wiersma1}, and the
other in the microwave response of a quasi-one dimensional system
\cite{Chabanov1}. Localization is expected to occur in the density
range given by the Ioffe-Regel condition, k\emph{l} $<$ 1, where k
is the local wave vector in the medium and \emph{l} the mean free
path for light scattering. In a dilute atomic vapor, the mean free
path is \emph{l} = $1/\rho\sigma$, where $\rho$ is the atom
density, and $\sigma$ the light scattering cross section
\cite{Sheng,LagTig}. Light localization has not yet been observed
in an atomic vapor, even though the density and temperature
regimes where it is expected to occur are technically accessible
using the techniques of ultracold atomic physics \cite{Metcalf}.

In addition to the intriguing possibility of strong light
localization in an atomic gas, research into other linear and
nonlinear effects in a multiple scattering environment is
relatively undeveloped \cite{ScullyZubairy}. There are also
potential ties to other areas of modern quantum optics research,
including the developments of quantum memories in the form of
polaritonic excitations \cite{FlLukin,LukinImam1} in strongly
scattering media. Manipulation of propagation and scattering in
ultracold atomic gases, through application of electromagnetically
induced transparency in various configurations
\cite{Harris,Liu1,Marangos}, could potentially lead to coherent
control of optical transport properties in such media. Similar
techniques have been applied to nonlinear optical phenomena
\cite{HarrisHau,Zhu} including four-wave mixing \cite{Braje1},
nonlinear optics at very low light levels \cite{Braje2}, and
applications of nonlinear optics with single photons to quantum
information processing \cite{LukinImam2}.

In this review we focus our discussion to near-resonance multiple
light scattering, in the weak localization regime, in ensembles
composed of ultracold atoms. We  review the various theoretical
and experimental results in this field, with some emphasis on our
theoretical development, but with due attention to the many recent
experimental achievements in this area of research. The main
physical observable is the coherent backscattering cone, which is
qualitatively characterized by an overall enhancement and angular
width.  We discuss the influence, in the coherent backscattering
effect, of ensemble size, optical depth, hyperfine and Zeeman
structure, and spectral detuning from resonance excitation. The
fascinating spatial variations in the angular distribution of
backscattering light produced by application of external static
magnetic fields of a few gauss are also reviewed. A dynamic area
of theoretical research, for which there are as yet only a few
experimental results, is the area of nonlinear optical effects in
coherent backscattering; we present an overview of these studies
as well. Of particular interest here is the time development of
the angular distribution and spectral profile of multiply
scattered near-resonance radiation.  For observations of the time
development of diffusely scattered light, experimental and
theoretical results show strong variations with light polarization
and detuning. In the backscattering direction, the theoretical
analysis of the time development of the scattered flux reveals a
variety of transient effects in the coherent backscattering
enhancement. Finally, we include in an appendix some details of
the theoretical developments which will be of interest to some
readers.

\section{Theoretical overview}
\setcounter{equation}{0}%
\subsection{Microscopic description}
Consider an atomic ensemble consisting of atoms separated on the
average by a distance larger than a typical radiative wavelength
$\lambda$. Let this ensemble scatter low intensity light, such
that the interaction process can be described properly by a
perturbation theory approach. Then in the Heisenberg formalism the
operator for the positive frequency component of the electric
field $\hat{\mathbf{E}}^{(+)}(\mathbf{r},t)$ at the point $\mathbf{r}$ and at the moment $t$, modified by the process of multiple scattering, can be expressed by the following series

\begin{eqnarray}
\hat{\mathbf{E}}^{(+)}(\mathbf{r},t)&=&\hat{\mathbf{E}}_0^{(+)}(\mathbf{r},t)\;+\;%
{\sum_{a}}\,\hat{\mathbf{E}}_{a}^{(+)}(\mathbf{r},t)%
\nonumber
\\%
&+&{\sum_{ab}}^\prime\,\hat{\mathbf{E}}_{ab}^{(+)}(\mathbf{r},t)\;+\;%
{\sum_{abc}}^\prime\,\hat{\mathbf{E}}_{abc}^{(+)}(\mathbf{r},t)\;+\;\ldots%
\nonumber
\label{2.1}%
\end{eqnarray}%

This expansion, written for the positive frequency component of
the electric field, shows how the different scattering orders,
starting from single via double and triple scattering, up to the
higher orders, subsequently contribute to the outgoing Heisenberg
operator. The indices $a$, $b$, $c$, etc. enumerate the atoms
participating in the scattering process. In double scattering
$a\neq b$, but in higher scattering orders recurrent scattering,
in which some of the indices coincide, can occur.

The expansion (\ref{2.1}) can be proved under the assumption that
at a microscopic level each randomly chosen and isolated atom of
the ensemble scatters light independently from its environment.
Then the series can be generated as an expansion of evolution
operators acting on the original "non-dressed" operator
$\hat{\mathbf{E}}_0^{(+)}(\mathbf{r},t)$ ignoring any interference
in interactions relating to different and well separated atoms.
This makes possible independent evaluation of each scattering
amplitude as well as the radiative correction of the excited state
atomic Green function. In the case of weak interactions, when the
incoming field does not noticeably modify the dynamics of the
atomic subsystem, the final operators of the positive/negative
field components $\hat{\mathbf{E}}^{(\pm)}(\mathbf{r},t)$ preserve
the canonical commutation relation between the non-perturbed
operators $\hat{\mathbf{E}}_0^{(\pm)}(\mathbf{r},t)$. Thus the
transformation (\ref{2.1}) is unitary and the whole series
reproduces correctly the microscopic behavior of the Heisenberg
field operator. This important property is based on the absence of
any losses of light apart from the scattering channel, see
\cite{KSKSH}.

As a pedagogical example, which will be used throughout our
discussion, let us show how a double scattering term can be
written in the case of successive scattering on atom "one" first
and on atom "two" second
\begin{eqnarray}
\hat{\mathbf{E}}_{12}^{(+)}(\mathbf{r},t) &=&\frac{1}{|\mathbf{%
r}-\mathbf{r}_{2}|r_{12}}\sum\limits_{m_{1},m_{1}^{\prime
}}\sum\limits_{m_{2},m_{2}^{\prime }}\,\sum\limits_{\nu}\sum\limits_{ij}%
\nonumber\\%
&&\sum\limits_{\mathbf{k},\mu}\,\left(\frac{2\pi\hbar\omega _{k}}{\mathcal{V}}\right)^{1/2}%
\frac{\omega _{2}^{2}}{c^{2}}\,\frac{\omega _{12}^{2}}{c^{2}}\,%
\nonumber \\%
&\times &\exp(-i\omega _{2}t+ik_{2}|\mathbf{r}-\mathbf{r}_{2}|+ik_{12}r_{12}+i\mathbf{k}%
\mathbf{r}_{1})%
\nonumber \\%
&\times&\mathbf{e}_{\mathbf{k}^{\prime }\nu }\,\hat{\alpha}_{\nu
i}^{(m_{2}^{\prime }m_{2})}(\omega _{12}-\mathbf{k}_{12}\mathbf{v}%
_{2})\,\delta _{ij}^{\perp }%
\nonumber\\%
&\times &\hat{\alpha}_{j\mu }^{(m_{1}^{\prime }m_{1})}%
(\omega-\mathbf{k}\mathbf{v}_{1})\,%
a_{\mathbf{k}\mu }%
\label{2.2}%
\end{eqnarray}%
It is assumed here that, before interaction, the light subsystem
is specified by the set of modes described by the wave vector
$\mathbf{k}$, frequency $\omega=\omega_{k}$ and the polarization
vector $\mathbf{e}_{\mathbf{k}\mu}$. Then $a_{\mathbf{k}\mu }$ is
an annihilation operator of the mode in the Schr\"{o}dinger
representation and $\cal{V}$ is the respective quantization
volume. For light scattered from atom "two" and propagating into
its radiation zone, similar parameters are respectively defined as
$\mathbf{k}'$, $\omega'$ and $\mathbf{e}_{\mathbf{k}'\nu}$. If
atoms are moving in space with velocities $\mathbf{v}_1$ and
$\mathbf{v}_2$ the input-output transformations of the light
frequency, as a result of successive quasi-elastic scattering
events, is a direct consequence of the combined action of Raman
processes and the Doppler effect.
\begin{eqnarray}
\omega _{12} &=&\omega \,-\,\omega _{m_{1}^{\prime }m_{1}}\,+\,%
(\mathbf{k} _{12}-\mathbf{k})\mathbf{v}_{1}%
\nonumber \\%
\omega'&=&\omega _{2}=\omega _{12}\,-\,%
\omega _{m_{2}^{\prime }m_{2}}\,+\,%
(\mathbf{k}_{2}-\mathbf{k}_{12})\mathbf{v}_{2}%
\label{2.3}%
\end{eqnarray}%
The light scattering is accompanied by Zeeman transitions
$|m_1\rangle\to|m_1'\rangle$ and $|m_2\rangle\to|m_2'\rangle$ in
the ground states of the first and the second atoms. The
intermediate and output wave vectors are given by
\begin{eqnarray}
\mathbf{k}_{12} &=&\frac{\omega _{12}}{c}\,%
\frac{\mathbf{r}_{2}-\mathbf{r}_{1}}{|\mathbf{r}_{2}-\mathbf{r}_{1}|}\,%
\approx \,\frac{\omega }{c}\,%
\frac{\mathbf{r}_{2}-\mathbf{r}_{1}}{|\mathbf{r}_{2}-\mathbf{r}_{1}|}%
\nonumber \\%
\mathbf{k}'&=&\mathbf{k}_{2}=\frac{\omega _{2}}{c}\,%
\frac{\mathbf{r}-\mathbf{r}_{2}}{|\mathbf{r}-\mathbf{r}_{2}|}\,\approx\,%
\frac{\omega }{c}\,%
\frac{\mathbf{r}-\mathbf{r}_{2}}{|\mathbf{r}-\mathbf{r}_{2}|}%
\label{2.4}%
\end{eqnarray}%
where the observation point $\mathbf{r}$ tends to infinity and the
approximated expressions, defined by the last equations and
ignoring all inelastic corrections, should be substituted in the
Doppler terms of (\ref{2.1}). By
$r_{12}=|\mathbf{r}_{2}-\mathbf{r}_{1}|$ we denoted the relative
distance between atom "one" and atom "two" which are located
respectively at spatial points $\mathbf{r}_1$ and $\mathbf{r}_2$.
Considering the reciprocal scattering path, i.e. scattering from
atom "two" first and atom "one" second it is only necessary to
transpose the indices $1\Leftrightarrow 2$ in the above
transformations, and the output frequency will obtain a different
magnitude $\omega_1$. But, as can be verified straightforwardly
for specific scattering channels such as the forward and backward
directions, the following equality is satisfied: $\omega ^{\prime
}=\omega _{2}=\omega _{1}$.

The most important characteristic contributing to (\ref{2.2}) is
the scattering tensor, which can be defined in operator form as
follows
\begin{eqnarray}
\hat{\alpha}_{ji}^{(m^{\prime }m)}(\omega)&=&%
-\sum_{n}\,|m^{\prime }\rangle \langle m|\,%
\frac{(d_{j})_{m^{\prime }n}\,%
(d_{i})_{nm}}{\hbar (\omega-\omega _{nm})\,+\,%
{\rm i}\hbar \gamma _{n}/2}%
\nonumber\\%
&\equiv& |m^{\prime}\rangle \langle m|\,%
{\alpha }_{ji}^{(m^{\prime }m)}(\omega)%
\label{2.5}%
\end{eqnarray}%
Here $d_i$ and $d_j$ are vector components of the transition
dipole moment between lower $|m\rangle$, $|m'\rangle$ and upper
$|n\rangle$ states, $\omega_{nm}$ is the transition frequency and
$\gamma_n$ is the natural relaxation rate of the upper state. As
long as $\gamma_n$ has a pure radiative nature the partial
transformation of the field operator in each scattering event,
described by the amplitude (\ref{2.5}), is unitary.

The $\delta ^{\perp }$-symbol is defined as
\begin{equation}
\delta _{ij}^{\bot}=\delta
_{ij}-\frac{k_{12i}\,k_{12j}}{k_{12}^{2}}%
\label{2.6}%
\end{equation}%
which guarantees that the light wave propagating between the
scatterers is transverse.

There are no additional physical ideas necessary to recover the
whole series (\ref{2.1}). Other terms contributing to this
expansion, such as $\hat{\mathbf{E}}_{abc}^{(+)}(\mathbf{r},t)$ for
triple scattering and the higher order terms, can be written
similar to (\ref{2.2}). For each scattering sequence
$a,b,c,\ldots$ the corresponding multiple amplitude will be a
subsequent product of scattering tensors and the photon Green
functions in vacuum, these being responsible for the light
propagation between the scatterers. Thus the entire series can be
recovered by following the simple combinatorial rules formulated
in the beginning of this section.

\subsection{Mesoscopic averaging and macroscopic description}

We restrict our discussion by application to the experiments
directed towards measurement of the first order interference or
correlation properties of light, which are described by the
following correlation function
\begin{equation}
D_{\mu\nu}^{(E)}(\mathbf{r},t;\mathbf{r}',t')\;=\;%
\langle\hat{E}_{\nu}^{(-)}(\mathbf{r}',t')\,%
\hat{E}_{\mu}^{(+)}(\mathbf{r},t)\rangle%
\label{2.7}%
\end{equation}%
Here the angle brackets denote statistical averaging over the
initial state of atoms and light. In the case of an ultracold
atomic gas that is not in a quantum degenerate phase, the location
of atoms can be visualized as the location of classical objects
distributed in a certain macroscopic volume. Then there are the
following important processes governing the averaging procedure.

First, for a light ray propagating in any direction, there is a
preferable coherent enhancement for its forward propagation. This
means that along any ray, for a short mesoscopic scale consisting
of a large number of atoms, there is only a slight attenuation of
the propagating wave. Such an attenuation comes from the events of
incoherent scattering, which have small but not negligible
probability. Then an important modification should be made for the
propagation of the light wave between any pair of neighboring
atoms, see example (\ref{2.2}), as well as for the incoming and
outgoing parts of the light path. The Green's function responsible
for the light propagation in a vacuum should be replaced by the
Fourier image of the retarded-type Green's function responsible
for the dispersion and intensity attenuation of the forward
propagating light in the bulk medium,
\begin{equation}
\delta _{ij}^{\perp}\,\frac{1}{r_{12}}\,%
\exp\left[{\rm i}k_{12}r_{12}\right]\;\to\;%
-\frac{1}{\hbar}D_{ij}^{(R)}(\mathbf{r}_{1},\mathbf{r}_{2},\omega_{12})%
\label{2.8}%
\end{equation}%
where $\omega_{12}$ is given by (\ref{2.3}). The complex conjugate
of Eq.(\ref{2.8}) transforms its right-hand side to the Fourier
components of the advanced-type Green function.

In a homogeneous and isotropic medium, the Green's function can be
introduced by direct modification of the exponential factor in the
left hand side of Eq.(\ref{2.8}) through absorption and refraction
indices of the medium. But in an inhomogeneous polarized atomic
gas it becomes a considerably more complex problem. Therefore, in
an appendix we briefly review the properties of the retarded
Green's function and show how it can be calculated in application
to the problem of light propagation through a polarized atomic
ensemble.

After averaging, the actual light wave in the sample can be
visualized as a set of unknown zigzag paths, whose vertices
consist of atoms scattering the light from the direction of
forward propagation. Any randomly chosen path contains a chain of
atoms located at the vertices and their number is just associated
with the scattering order. Each chain of atomic scatterers makes a
partial contribution to the formation of the outgoing wave similar
to how it is described by expressions (\ref{2.1}) for the
Heisenberg operators. The important difference is that, as a
result of the mesoscopic averaging, the series {\it converges
rapidly} and only the multiple scattering of the {\it low orders}
contribute significantly to the formation of the correlation
function (\ref{2.7}).

Secondly, it is remarkable that the coherence is not completely
lost for scattering in the non-forward direction. For the light
emerging from the sample in the backward direction the
interference of multiple amplitudes for any selected chain of
scatterers survives statistical averaging. This is known as the
coherent backscattering (CBS) effect, which is closely related to
weak localization of light. Comparing the expression (\ref{2.2})
with a similar one written for the reciprocal
$\hat{\mathbf{E}}_{21}^{(+)}(\mathbf{r},t)$ term, the CBS effect
as well as the criteria of its observation can be clearly seen. If
scattered light is detected at any random angle, such as
$\mathbf{k}^{\prime }+\mathbf{k}\neq 0$, the interference
contribution becomes quite sensitive to the atoms locations. In a
sample consisting of many atoms the interference, being averaged
over all possible combinations of atomic pairs, will be negligible
comparing with ladder (non-interference) term. But this would not
be the case in observation of the scattering in the backward or
near-backward direction, where $\mathbf{k}^{\prime }+\mathbf{k}\to
0$. Weak oscillations caused by atomic motion or Raman type
scattering still survive. But these also become negligible in the
case of elastic scattering on cold and slowly moving atoms.

\subsection{Observable characteristics in coherent backscattering}

Normally the relevant quantity for discussing the scattering
process is a differential cross-section, which is defined as the
normalized flux of the scattered light emerging the sample in the
observation direction. In terms of the correlation function the
cross section is given by expression (\ref{2.7}) considered at
coincident spatial and time arguments:
$\mathbf{r}=\mathbf{r}',\;t=t'$. In this section we illustrate by
certain physical examples that the spectrally-sensitive and
time-dependent measurements of the light correlation function
(\ref{2.7}) provide us further information than the measurement of
the cross-section only. We concentrate ourselves on the time
dependence of the correlation function, which corresponds to the
following observation schemes.

First, the excitation of the ensemble can be initiated by a
coherent light pulse. In this case the original correlation
function will be factorized in the product
\begin{equation}
{\cal D}_{\mu\nu}^{(E)}(\mathbf{r},t;\mathbf{r}',t')\;=\;%
{\cal E}_{\nu}^{(-)}(\mathbf{r}',t')\,%
{\cal E}_{\mu}^{(+)}(\mathbf{r},t)%
\label{3.1}%
\end{equation}%
where ${\cal E}_{\mu}^{(+)}(\mathbf{r},t)$ is a coherent field
component of the laser light pulse. Then the scattered response
(\ref{2.7}), whose shape is a distorted copy of the original pulse
profile, can provide us with comparative information about how
this response is sensitive to the effects of single and higher
orders of the multiple scattering. The appropriate analysis of the
scattered pulse can be made by the methods of time-dependent
spectroscopy.

Second, atomic motion, which always exists in a realistic sample,
leads to a random low-frequency modulation of the scattering terms
because of the Doppler effect. As clearly seen in the example of
double scattering (\ref{2.2}), such a modulation is described by
velocities $\mathbf{v}_1$ and $\mathbf{v}_2$ considered as
stochastic parameters in the frequency transformation (\ref{2.3}).
Then the probe of the sample with a monochromatic coherent wave of
frequency $\omega$ will be modified in response as a
non-monochromatic scattered wave with the output correlation
function (\ref{2.7}) decaying as a function of $t-t'$. Taken at
coincident spatial arguments ${\mathbf r}={\mathbf r}'$ and for a
point-like photodetector, the correlation function can be
expressed in the form
\begin{equation}
\frac{c}{2\pi}\,D_{\mu\nu}^{(E)}(\mathbf{r},t;\mathbf{r},t')\;=\;%
{\mathrm e}^{-i\omega_{R}\tau}\,I_{\mu\nu}(\tau)%
\label{3.2}%
\end{equation}%
where $\omega_{R}$ denotes the carrier frequency of the scattered
light, which in general can be shifted from the input frequency
$\omega$ because of the inelastic Raman effect. The outgoing
intensity in any selected polarization channel is described by the
Pointing vector $I_{\mu\nu}{\mathbf k}'/k'$ (for $\mu=\nu$) and is
given by the correlation function considered at coincident times
$t=t'$. The dependence on $\tau=t'-t$ in the right hand side comes
from the spectral distribution of the scattered modes. The Fourier
transform
\begin{equation}%
I(\omega)=\sum_{\mu=1,2}\,\int_{-\infty}^{\infty}%
d\tau {\mathrm e}^{i(\omega-\omega_R)\tau}\, I_{\mu\mu}(\tau)%
\label{3.3}%
\end{equation}%
gives us the spectral distribution of the scattered intensity in
the vicinity of the Raman frequency. As we see, the knowledge of
the spectral distribution (\ref{3.3}) provides us with quite
important information about the velocity distribution and possible
correlations existing in an atomic ensemble confined with a
magneto-optic trap. The corresponding spectral selection can be
done by heterodyne detection and the light beating spectroscopy
method \cite{Lightbeating}.

We conclude the theoretical overview by the following remark. If
an atomic sample were excited with monochromatic mode and the
effect of atomic motion were neglected, then all the
characteristics of the scattered light would be completely
described inside the cross-section formalism. The spectrally
sensitive and time-dependent analysis gives us more access to the
important physical information concerning the light propagation
and internal dynamics of the atomic gas in the magneto optic trap.

\section{Experimental overview}
\setcounter{equation}{0}%
\subsection{Apparatus}
In this section we give a broad overview of an experimental
apparatus used to measure light scattering in an ensemble of atoms
consisting of an ultracold, dilute gas of atomic $^{85}$Rb
confined in a magneto-optical trap (MOT). The description here
refers in particular to that of Ref. \cite{KSKSH}; the physics of
the process is such that the described approach is quite general.
In the present case, the MOT operates on the $5s
^{2}S_{1/2}\;F_0=3 \rightarrow 5p ^{2}P_{3/2}\;F=4$ hyperfine
transition and produces a nearly Gaussian cloud of approximately
10$^{8}$ atoms at a temperature $\sim$100 $\mu$K. The peak density
at the center of the trap is $\sim $3 x 10$^{10}$ cm$^{-3}$. The
Gaussian radius of the sample is $r_{0}$ $\sim$ 1mm, determined by
fluorescence imaging. Measurement of the spectral variation of the
transmitted light gives a peak optical depth, through the center
of the trap, of $b_{0}$ = 6 - 8. For a Gaussian atom distribution
in the trap, the weak-field optical depth, on resonance and
through the center of the trap, is given by
$b_{0}=\sqrt{2\pi}n_0\sigma_{0}r_{0}$, where $n_0$ is the peak
trap density and $\sigma_0$ is the resonance cross section, see
section \ref{S4.1}.

Note that for an isolated transition the near
resonance cross section for light scattering $\sigma$ and the
respective optical depth $b$ vary with probe frequency such that
\begin{equation}
b = \frac{b_{0}}{1 + (2\Delta/\gamma)^{2}}, \label{b}
\end{equation}
where $\Delta = \omega_{L} - \omega_{0}$, and $\omega_{L}$ is the
probe frequency, while $\omega_{0}$ is (in the present case) the
$F_0=3 \rightarrow F=4$ resonance frequency.

Separate lasers are used to provide the trapping and probe light.
In both cases, a continuous wave diode laser having a bandwidth
$\sim$ 1 MHz is used. A full description of the master-slave laser
system and vacuum hardware can be found in \cite{KSKSH}. The laser
intensity for both the trapping and probe light is modulated with
an acousto-optic modulator (AOM) - used as an optical switch -
which generates nearly rectangular pulses of adjustable duration.
The 20 dB response is limited by the AOM to about 60 ns. The laser
light is subsequently coupled into a single mode fiber optic
patchcord. The combination of the AOM switching and fiber coupling
results in an $\sim$65 dB attenuation of the laser light when
switched off. A weak probe laser is tuned in a range of several
$\gamma$ around the trapping transition. The probe laser is
linearly polarized in the vertical direction. The probe beam is
directed into the MOT as shown in Figure 1.

\begin{figure}[tp]
{\includegraphics{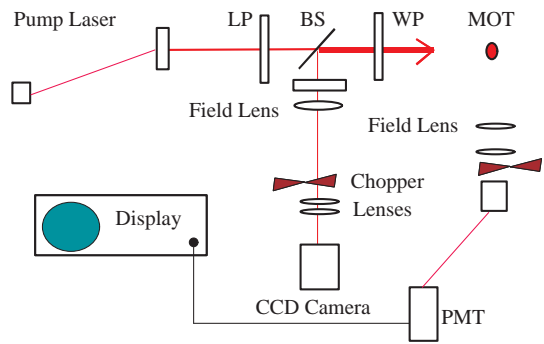}}
\caption{A schematic diagram of a typical experimental
arrangement. Shown in the figure is a magneto optical trap (MOT),
linear polarizers (LP) which select the detected polarization
channels, a beam splitter (BS), wave plates (WP) and a
photomultiplier tube (PMT) to detect the fluorescence signals for
time dependent measurements.  For continuous wave measurements, a
charge-coupled device (CCD) camera is used.  The generic display
is a multichannel scalar for time dependent measurements.}
\label{Fig.1}
\end{figure}

Depending on the experiment to be performed, light scattered by
the atoms is detected either in the backward direction on a liquid
nitrogen cooled charge-coupled device (CCD) camera or in a
direction at some other angle relative to the incident probe beam
using a photomultiplier tube (PMT).

\subsection{Coherent Backscattering Measurements}
For measurements of the CBS cone, great care must be taken to
suppress multiple and back reflections from optics in the
detection optical path. A major source of unwanted back-scattered
light is from the vacuum viewports on the MOT chamber. Windows are
typically wedged and V-type anti-reflection (AR) coated for 780 nm
on the probe laser entrance and exit ports and on the CCD camera.
The AR coating characteristically results in less than 0.25 \%
reflectivity at 780 nm. Additionally, the entrance port window can
be mounted on a ultrahigh vacuum bellows, allowing redirection of
unwanted reflections away from the detector.

After exiting the fiber, the probe beam was expanded and
collimated by a beam expander to a 1/e$^{2}$ diameter of about 8
mm. The polarization of the resulting beam was selected and then
the beam passed through a nonpolarizing and wedged beam splitter
that transmits approximately half of the laser power to the atomic
sample. The backscattered radiation is directed by the same beam
splitter to a field lens of 45 cm focal length, which condenses
the light on the focal plane of the CCD camera. The diffraction
limited spatial resolution was about 100 $\mu$rad, while the
polarization analyzing power is greater than 2000 at 780 nm. Any
one of the four polarization channels that are customarily studied
in coherent backscattering can be selected by inserting or
removing the quarter wave plate, and adjusting the linear
polarization analyzer located before the field lens.

\subsection{Time-Dependent Scattering Measurements}
For measurements of time-dependent light scattering, light signals
are detected in a direction away from the coherent beam. In a
typical geometry, as shown in Figure 1, detection could be in a
direction orthogonal to the probe laser propagation and
polarization directions. For example, in Ref. \cite{Balik}, the
light was collected in an effective solid angle of about 0.35
mrad, and refocussed to match the numerical aperture of a 400
$\mu$m multimode fiber. A linear polarization analyzer is placed
between the MOT and the field lens to collect signals in
orthogonal linear polarization channels, which we label as
parallel ($\|$) and perpendicular ($\bot$). The differential
polarization response is calibrated against the known linear
polarization direction of the probe laser, and the measured 20 \%
difference in polarization sensitivity is used to correct the
signals taken in the two polarization channels. The fiber output
is coupled through a 780 nm (5 nm spectral width) interference
filter to a near-infrared sensitive GaAs-cathode photomultiplier
tube. The PMT output is amplified and directed to a discriminator
and multichannel scalar, which serves to time sort and accumulate
the data into 5 ns bins. A precision pulse generator is used to
control the timing of the MOT and probe lasers and for triggering
the multichannel scalar.

Finally, we point out that in this type of experiment the
quantitative results obtained depend on the relative diameter of
the pumping beam and the sample size.  The main effect is in the
contribution of single scattering in comparison with the multiple
scattering signals.  For a pump beam large in comparison with the
samples, there is significant single scattering from the
relatively low density periphery, while for a narrow probe beam a
larger portion of the signal is due to scattering from the denser
regions of the sample.

\section{CBS observation in an ultracold atomic gas in the continuous wave regime}
\setcounter{equation}{0}%

\subsection{The enhancement factor and cone shape} \label{S4.1}

\begin{figure}[tp]
{\includegraphics{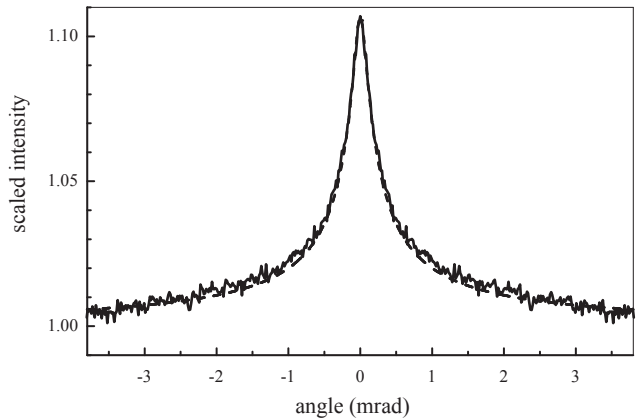}}
\caption{A coherent backscattering cone in the helicity
non-preserving channel associated with the $F_0 = 3 \rightarrow F
= 4$ transition in ultracold atomic $^{85}$Rb. Reprint of Fig. 3
from Ref. \cite{LTBMMK}; copyright 1999 by the American Physical
Society.}
\label{Fig.F1}%
\end{figure}

First observation of the CBS effect in an ultracold atomic gas was
reported by Labeyrie \emph{et al.} in Ref. \cite{LTBMMK}. In
Figure \ref{Fig.F1} we reproduce the experimental graph from that
paper showing the cone feature in the spatial profile of the
backscattered light. The experiment was done with atoms of
${}^{85}$Rb and the scattered light was observed as a response to
a probe laser tuned near-resonance with the closed hyperfine
transition $F_0=3\to F=4$ of the rubidium $D_2$ line. The rubidium
atoms form a convenient sample for measurements and the dependence
of Figure \ref{Fig.F1} shows a typical behavior of the CBS cone
from an ensemble of these atoms. The measurements shown in Figure
\ref{Fig.F1} \cite{LTBMMK} were made with a circular polarized cw
monochromatic probe laser, and the scattered light was detected in
the channel with orthogonal helicity, which relates to the
Rayleigh process for the single scattering event.

\begin{figure}[tp]
{\includegraphics{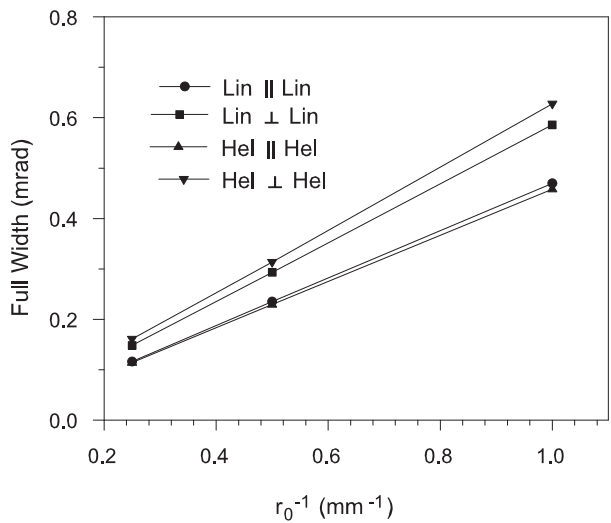}}
\caption{Dependence on the inverse sample size of the full width
at half maximum of the coherent backscattering cone for various
polarization channels. The peak optical depth is fixed at five for
these calculations.  Calculations refer to Monte-Carlo simulations
of light scattering on the $F_0=3 \rightarrow F = 4$ resonance
transition in ultracold atomic $^{85}$Rb.}
\label{Fig.F2}%
\end{figure}%

The graph shown in Figure \ref{Fig.F1} gives us the following two
important parameters of the CBS process. The main informative
parameter is the so-called enhancement factor, which is defined as
\begin{equation}
\alpha=1 + \frac{I_C}{I_L+I_S}%
\label{F1}\end{equation}%
and shows the maximum enhancement of the backscattered intensity.
As was discussed in the theoretical overview, the additional
intensity in the scattered light is a result of constructive
interference between direct and reciprocal scattering paths and is
described by the cross terms $I_C$ in the numerator of (\ref{F1}).
The denominator consists of the single scattering contribution
$I_S$ and non-interfering ladder contributions $I_L$ of the second
and higher orders of multiple scattering. As is clear from the
structure of Eq. (\ref{F1}), for classical type dipole scatterers
with only Rayleigh scattering channels, the enhancement factor
$\alpha$ should approach $2$ if the optical depth $b_0$ tends to
infinity. But it is also clearly seen that this is not the case
for the experimental graph shown in Figure \ref{Fig.F1}.

As was pointed out later in \cite{JMKMD} the relatively small
magnitude of the enhancement factor is a direct consequence of
multi-Zeeman-level atomic structure. The physics of the process
was reiterated within an analytical microscopic theory developed
by M\"{u}ller, \emph{et al.} \cite{Muller}. The enhancement of a
factor of two can be achieved only if direct and reciprocal
scattering amplitudes describe the time reversal processes.
Otherwise, for weak field scattering, the interference always
leads to an enhancement less than a factor of two. Recovery of the
factor of two in the enhancement factor was clearly demonstrated
in experiments on atomic Sr by Bidel, \emph{et al.}
\cite{Strontium1}; in this case scattering is on a $J_0=0
\rightarrow J = 1 $ transition. The differences between the Sr and
Rb cases, including the role of nonzero atom velocity, was
emphasized in Wilkowski, \emph{et al.} \cite{Wilkowski1}. Finally,
we point out that the interferences in coherent backscattering can
be destructive in special scattering channels. Below we discuss
such an example of destructively interfering channels, where the
enhancement factor is less than unity.

The second important characteristic of the cone shape is its
angular width. For simple evaluation and as a pedagogical model,
one can imagine a semi-infinite homogeneous medium and apply the
diffusion approach to estimate the statistical distribution of
spatial separations of the first and last scatterers in a
scattering chain. It is only the location of these scatterers that
determines the phase of the interference in the cross term. This
straightforwardly leads to the following estimation of the cone
angle
\begin{equation}
\theta_{\rm CBS}'\sim \frac{1}{kl_0}%
\label{F2}\end{equation}%
where $l_0=1/n_0\sigma_0$ is the free path for a resonance photon
migrating in the sample, $n_0$ is the density of scatterers and
$\sigma_0$ is the resonance cross section. However the application
of this estimation to real experiments with atomic scatterers
confined in a MOT fails and quantitatively disagrees with
observable data.

As was shown by precise Monte-Carlo modelling in \cite{KSKSH}, and
was also discussed in \cite{LDMMK}, the cone angular width does
not depend on the diffusion length in the case of an atomic cloud
with a Gaussian-type density distribution. For the density
distribution
\begin{equation}%
n(r)=n_0\,\exp(-\frac{r^2}{2r_0^2})%
\label{F3}%
\end{equation}%
where $n_0$ is a peak density in the middle of the cloud and $r_0$
is the radius of the cloud, the relevant estimation of cone angle
is given by
\begin{equation}
\theta_{\rm CBS}\sim \frac{1}{kr_0}%
\label{F4}\end{equation}%
This is illustrated by the respective dependencies shown in Figure
\ref{Fig.F2}, which shows the linear dependence of the cone width
on inverse sample size $r_0^{-1}$ in various polarization channels
of the $F_0=3\to F=4$ transition of ${}^{85}$Rb. Since for the
Gaussian cloud the optical depth is given by
$b_0=\sqrt{2\pi}\sigma_0n_0r_0$ we see that the estimations
(\ref{F2}) and (\ref{F4}) differ by a factor of $b_0$ and
expression (\ref{F2}) gives a larger cone angle width than is
actually measured.

The CBS images in the plane orthogonal to the incident and
backscattered directions are different for different polarization
channels. Normally the polarization channels are discussed for
circular and linear input and output polarizations. Circular
polarization is normally defined in terms of helicities (hel) with
respect to the frame of wave propagation, while linear
polarization directions (lin) are defined with respect to a
laboratory frame. The scattered light is detected either in the
same polarization channel as the input light or in an orthogonal
polarization channel. There is a spatial asymmetry in the linear
polarization channel. This includes a larger cone width in the
vertical direction for the lin $\parallel$ lin channel with the
lines of asymmetry along the bisectors of the detected linear
polarization directions in the lin $\perp$ lin channel. The
details of the features in the cone shape are discussed in Ref.
\cite {KSBHKS}.

\subsection{The influence of hyperfine structure}\label{S4.2}

\begin{figure}[tp]
{\includegraphics{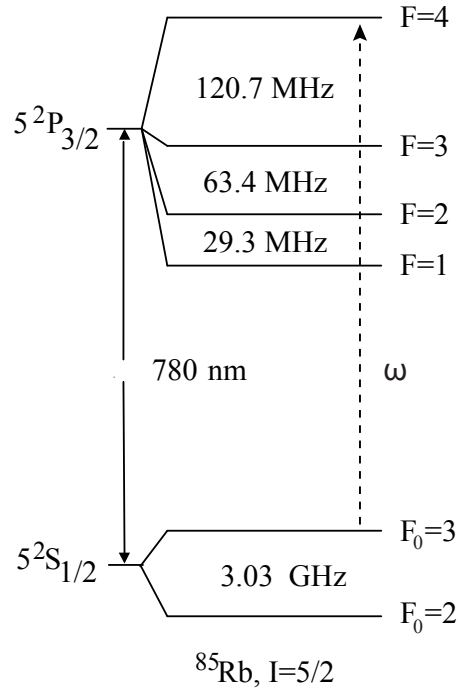}}
\caption{Hyperfine energy levels of relevant transitions in atomic
${}^{85}$Rb.}
\label{Fig.F3}%
\end{figure}

In initial studies of coherent backscattering in atomic samples,
it was assumed that for near-resonance scattering the hyperfine
structure of the excited atomic state is unimportant (other than
for the degeneracies of the transitions under consideration).
Particularly for the $F_0=3\to F=4$ "closed" transition in
${}^{85}$Rb the nearest hyperfine satellite $F_0=3\to F=3$ is
located at $-120$ MHz, see Figure \ref{Fig.F3}. This is about $20$
times the natural line width $\gamma\sim 5.9$ MHz and argues for
the unimportance of the off-resonant transitions. However as was
predicted in \cite{KSKSH} there is an asymmetry in the CBS
enhancement for the spectral scanning near the resonance, this
being caused by the interference among all the hyperfine
transitions. The asymmetric shape is more clearly seen in the case
of circular polarization. This indicates the non-trivial spectral
behavior of the Raman-type cross/interference terms and the
Raman-type ladder terms near the resonance.

\begin{figure}[hb]
\centering
{\includegraphics[width=200pt,height=200pt]{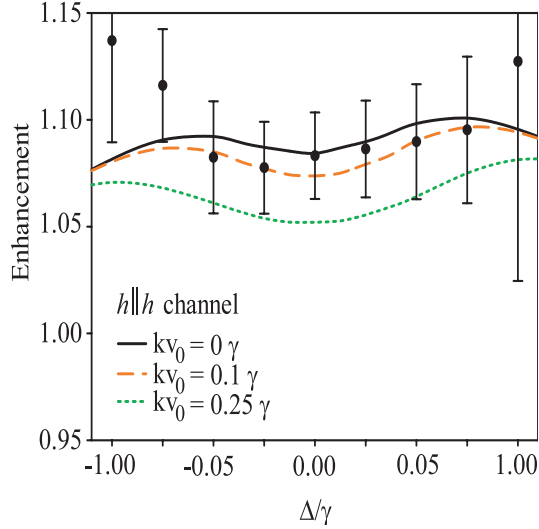}}
\caption{Comparison of experimental and theoretical enhancement
spectra in the helicity-preserving polarization channel associated
with an $F_0=3 \rightarrow F = 4$ resonance transition.
Theoretical spectra show modification by Doppler broadening, which
is varied from $kv_0=0$ to $kv_0=0.25\gamma$, in an ensemble of
${}^{85}$Rb atoms having a peak density of $n_0=1.6\times 10^{10}$
cm$^{-3}$ and a Gaussian radius $r_0=1$ mm.}
\label{Fig.F4}%
\end{figure}

The experimental verification of this effect was made in
\cite{KSLKSBH,LDMMK1} and in Figure \ref{Fig.F4} we reproduce the
illustrative graph from \cite{KSLKSBH} for the helicity preserving
scattering channel. In the calculations the influence of possible
heating effects, where the Doppler broadening $kv_0$
($v_0=\sqrt{2k_BT/m}$ is the most probable velocity in the atomic
ensemble) was varied from $0$ to $0.25\gamma$. The heating
mechanism makes the enhancement weaker but the spectral shape
broader. Comparing the results one can see that there is a
qualitative agreement between theoretical calculations and
experimental data.  However they differ quantitatively and the
enhancement, observed in the experiment in the wings, is even
larger than its theoretical prediction. As was mentioned in
\cite{KSLKSBH} the main possible reason of such a contradiction
between experiment and theory is in the optical pumping process,
which tends to orient atomic spins along the probe beam. For 100\%
orientation this effect can increase the enhancement up to maximum
value of two. The enhancement of a factor of two would be
achievable because for spin oriented ensemble there is no the
single scattering contribution. Then there would be only two
constructively interfering amplitudes if the double scattering
channel could be isolated. This effects is similar to the
enhancement factor behavior in a strong magnetic field, as
discussed in the following section.

We conclude this section by emphasizing that there are also
significant variations in the enhancement factor even for resonant
scattering on different $F_0 \rightarrow F$ transitions.  This is
due primarily to the different degeneracies associated with the
transitions.  Experimental confirmation of these variations, and
comparisons with Monte-Carlo and model calculations have recently
been reported by Wilkowski, \emph{et al.} \cite{FrenchHyperfine}.

\subsection{Influence of an external magnetic field}

The influence of an external magnetic field on the multiple
scattering in optically dense atomic ensembles was studied in
Refs. \cite{LMK,LMMSDK,SLJDKM}. The unique conditions of ultracold
atomic ensembles, where there is negligible inhomogeneous Doppler
broadening, make possible the spectroscopic manipulation with a
magnetic field for the Zeeman splitting of the atomic levels at
the level of the natural line width. In Ref. \cite{LMK}, which has
no direct relation to the CBS phenomenon, the Faraday rotation of
quasi-resonant light in an optically thick cloud of laser cooled
rubidium atoms was experimentally studied. Measurements yield a
large Verdet constant in the range $200000^0$/T/mm and a maximal
polarization rotation of $150^0$. The Faraday effect was initiated

by the Zeeman splitting in the ground and in the upper state,
which led to differences in the refraction indices for
$\sigma_{+}$ and $\sigma_{-}$ polarization of the probe light. It
is just the absence of the Doppler effect and the possibility to
scan the sample near the natural resonance line which permits such
a huge Verdet constant to be obtained.

As a next step it was recognized that a weak magnetic field, with
Zeeman splitting comparable with the natural line width, should
modify the CBS process itself. The presence of the magnetic field
manifests itself in the scatterer dipole response to the electric
field in a manner similar to the Hanle effect in a fluorescence
geometry. As was shown in experiment \cite{LMMSDK} the polar shape
of the CBS cone in the linear polarization scattering channel
follows the magnitude and direction of the magnetic field vector.
In the theoretical discussion of Ref. \cite{LMMSDK} an attempt was
made to classify the influence of a magnetic field in terms of the
well known Faraday, Cotton-Mouton and Hanle effects. But as is
clear in multiple scattering, and particularly in the multiple
scattering regime of CBS, there is only a convenient analogy with
these basic optical processes.

\begin{figure}[tp]
{\includegraphics{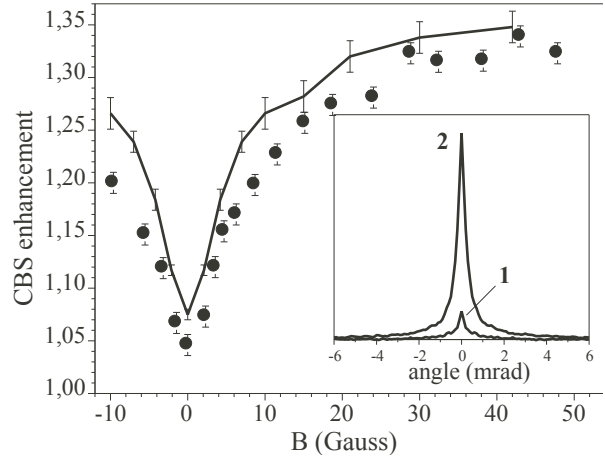}}
\caption{CBS enhancement factor of light for a cold Rb atomic
cloud, measured in the parallel helicity channel hel $\parallel$
hel, as a function of the transverse magnetic field strength $B$;
the graph shows an increase in the CBS enhancement for non-zero
magnetic field. The solid line is the result of a Monte-Carlo
simulation. The inset shows profiles of the CBS cones when $B = 0$
(1) and $B = 43$ G (2). Reprint of Fig. 2 from Ref. \cite{SLJDKM};
copyright 2004 by the American Physical Society.}
\label{Fig.F5}%
\end{figure}

The remarkable manifestation of a magnetic field was recently
observed in Ref. \cite{SLJDKM}. There it was observed that the
enhancement factor can be increased with magnetic field up to its
maximal value of two. This unusual behavior appears due to lifting
of degeneracy in the helicity scattering channel for the
spectrally selected Zeeman hyperfine transition $F_0=3,M_0=3\to
F=4,M=4$ of ${}^{85}$Rb, which can be done by applying a rather
strong external magnetic field. For this transition there is no
Raman-type scattering in the single scattering response. If only
double scattering dominates in the helicity preserving scattering
response there should be a factor of two enhancement of the
scattered intensity. In the Figure \ref{Fig.F5} we reproduced the
basic graph of \cite{SLJDKM}, showing the experimental
verification of this effect. Let us also point out that there is a
direct analogy of the behavior of the enhancement factor in a
magnetic field with its spectral behavior in a spin-oriented
ensemble predicted in \cite{KSLKSBH}.

\subsection{The CBS process in the saturation regime}

Recent experiments by Chaneliere, \emph{et al.} \cite{Chaneliere}
on the resonance transition in atomic strontium and by Balik,
\emph{et al.} \cite{Balik} in rubidium, have shown that a
significant reduction in the coherent backscattering enhancement
can occur with increasing intensity of the probe laser. Both
non-linear effects and additional inelastic scattering components
can contribute to this reduction. Theoretical and model studies
have shown similar qualitative effects
\cite{Wellens1,Shatokhin,Wellens2}, although they have yet to be
quantitatively compared with experiment.

In the strontium experiment, a MOT containing 7$\times$10$^{7}$
atoms at a temperature of $\sim$1mK and with a Gaussian radius of
$\sim$0.7 mm was illuminated with near resonant light. The
ensemble of cold atoms had a peak optical depth of $b_{0}$ = 3.5
and $\emph{kl}_0=10^{4}$, resulting in a regime of weak
localization. In the absence of MOT light and magnetic field
gradients, a resonant probe beam illuminated the sample for a
variable period of 5 to 70 $\mu$s. The probe pulse duration was
adjusted so as to keep the total number of scattered photons below
400 for all intensities investigated, thereby minimizing
mechanical effects of the light on the cold atom sample. A CBS
cone was then recorded in the helicity preserving channel (hel
$\|$ hel), where single scattering is negligible. Keeping the
shape of the CBS beam constant, but treating the width and
enhancement factors as free parameters, a CBS cone enhancement was
extracted. Results for on resonance scattering and a detuning of
$\delta=\Gamma/2$ were measured as a function of saturation
parameter $s=I/I_{s}$, where $I_{s}$=42 mW/cm$^{2}$. The results
indicate that saturation is in part responsible for the magnitude
of the cone enhancement, but saturation alone does not fully
describe the CBS process. Detuning also plays a role because the
ratio of inelastic to elastic scattering is detuning dependent. It
is the inelastic component which degrades reciprocity and causes
the enhancement factor to decrease. As the field increases, the
inelastic nature of the light scattered also increases yielding a
decreasing cone enhancement factor.

In the rubidium experiment, a nearly spherical MOT containing
$\sim4\times10^{8}$ at a peak density of $1.6\times10^{8}$
atoms-cm$^{-3}$ was illuminated with a probe beam for time T and
the CBS cone was recorded on a liquid nitrogen cooled CCD camera
as described in Section 3. Two polarization channels, lin $\bot$
lin and hel $\|$ hel were investigated. The intensity of the
backscattered light was detected and the enhancement factor of the
CBS cone measured as a function of probe intensity, again
expressed in terms of the saturation parameter s, where $I_{s}$ in
this case is the on-resonance saturation intensity of about 1.6
mW/cm$^{2}$. Data taken at s=0.08 provided the baseline for the
``weak" field cone measurements. The baseline saturation parameter
and exposure time were selected to minimize mechanical action of
the CBS laser on the cold atom sample as in the Sr experiment.
With s = 0.08, a probe exposure time of T = 0.25 ms satisfied this
condition.

Measurements up to a saturation parameter of 9 in the lin $\bot$
lin channel were made. For optical depths larger than unity (as is
the case here), one expects the width of the cone to be determined
primarily by the spatial distribution of atoms in the MOT and not
be very sensitive to the saturation parameter. Indeed, little
change in the angular width of the cone was observed, in agreement
with theory and the experimental results of Ref.
\cite{Chaneliere}. The cone enhancement factor $\alpha$, however,
was observed to decrease substantially in the strong field regime.
Cone images for the lin $\bot$ lin channel are shown in Figure 7.
In this data, the product sT was constant and the total spatially
integrated intensity was the same in all four images.

\begin{figure}[tp]
{\includegraphics{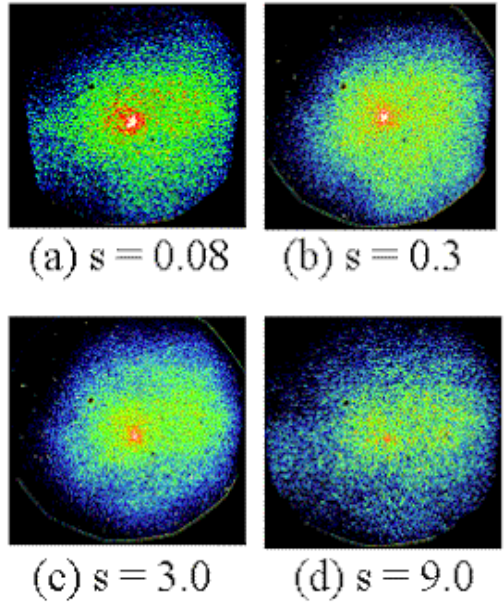}}
\caption{Dependence on the saturation parameter s in the lin
$\bot$ lin polarization channel, of CCD images of the intensity
field in the nearly backscattering direction. The data correspond
to the $F_0 = 3 \to F = 4$ hyperfine resonance in $^{85}$Rb.}
\label{Fig.2}%
\end{figure}%

The data, which spans a saturation parameter range of more than
100, shows a clear reduction in the contrast of the cone with
increasing intensity. This behavior is shown quantitatively in
Figure 8, where it is seen that the enhancement $\alpha$ decreases
with increasing saturation parameter over the data range explored.
As previously discussed, a similar monotonic decrease in $\alpha$
has been observed in experiments on the singlet resonance
transition in ultracold Sr \cite{Chaneliere}. However, the
percentage decrease in $\alpha$ with increasing s was observed to
be much larger in that case than in the Rb experiments.

\begin{figure}[tp]
{\includegraphics{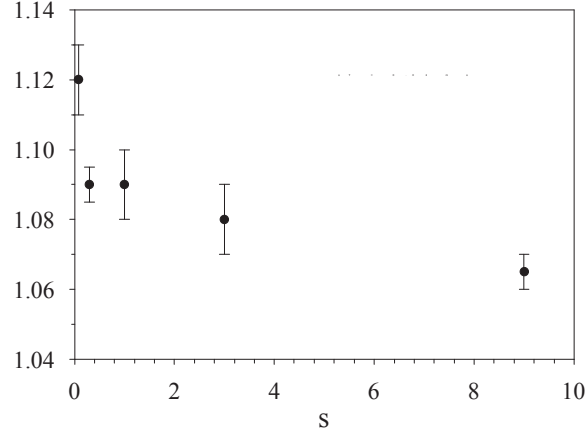}}
\caption{Dependence of the coherent backscattering enhancement on
the saturation parameter s in the lin $\bot$ lin polarization
channel. The data correspond to the $F_0 = 3 \to F = 4$ hyperfine
resonance in $^{85}$Rb.}
\label{Fig.2}%
\end{figure}%

\begin{figure}[tp]
{\includegraphics{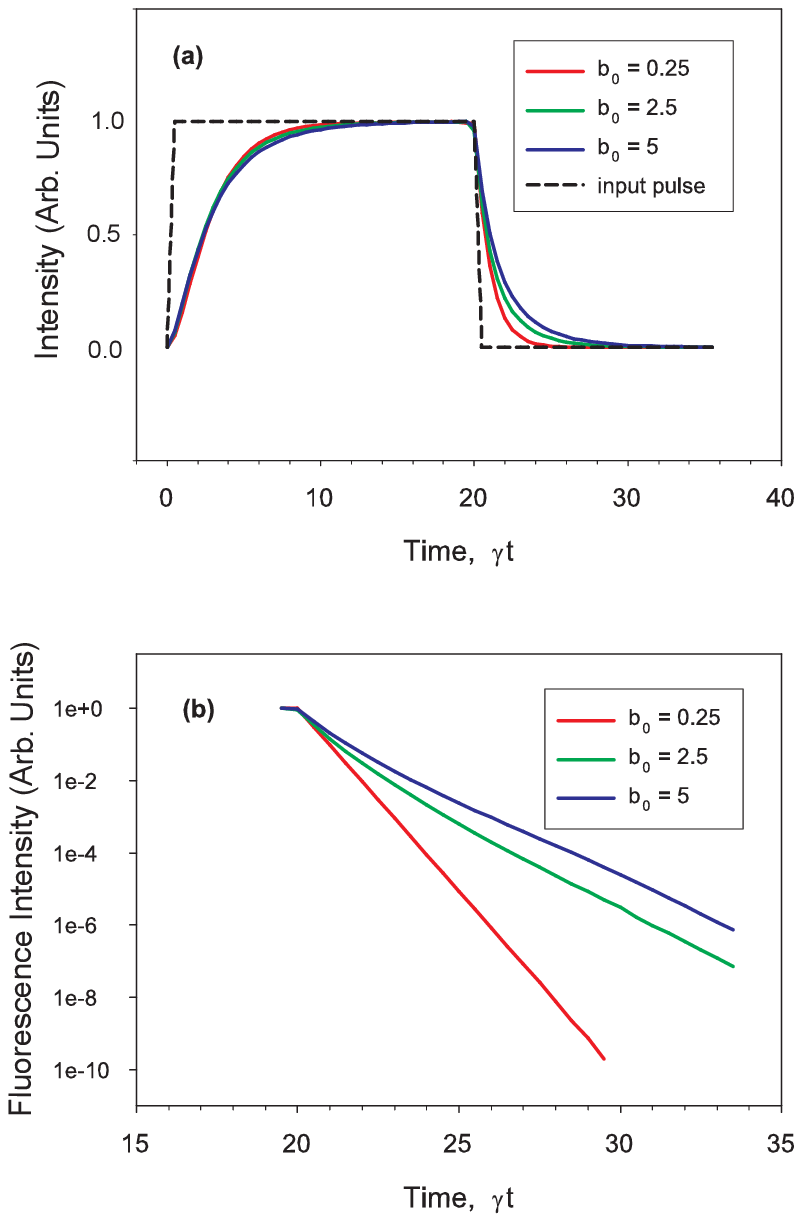}}
\caption{(a) The intensity response for light scattered in the
backward direction from an atomic sample of ${}^{85}$Rb probed by
a pulse with a duration of $\tau_0=20\,\gamma^{-1}$ tuned to
resonance with the $F_0=3\to F=4$ hyperfine transition for
different optical thicknesses $b_0$. (b) Fluorescence decay after
the excitation pulse is switched off.}
\label{Fig.1}%
\end{figure}%

Measurements of $\alpha$ in the hel $\|$ hel channel (not shown)
in the range 0.04 $\leq$ s $\leq$ 1.0 stay relatively constant -
in sharp contrast to the results for the lin $\bot$ lin channel
and to the results of Chaneliere, \emph{et al.} \cite{Chaneliere}.
The approximate constancy of $\alpha$ is a surprising result and
suggests that optical pumping of the Zeeman sublevels in the
ground level may play an important role in the observed
quantitative results. Several other possible physical mechanisms
explaining the data have been suggested in \cite{Balik} but await
confirmation by detailed calculations.

\section{Time-dependent and spectral analysis of the scattered light}
\setcounter{equation}{0}%

\subsection{Time dependent spectroscopy in a backscattering geometry}\label{3A}
Consider excitation of an atomic ensemble with a coherent light
pulse, when the correlation function can be factorized in the form
(\ref{3.1}). Let the field amplitude of ${\cal
E}_{\mu}^{(+)}(\mathbf{r},t)$ have a rectangular pulse profile. It
is convenient to scale the duration of the pulse $\tau_0$ by the
natural atomic lifetime $\gamma^{-1}$. In accordance with
(\ref{2.1}) and with the usual restrictions of the rotating wave
approximation, the rectangular pulse profile should be expanded
over the set of incoming modes with the following amplitude
\begin{equation}%
\alpha_{\mathbf k\mu}\propto %
\frac{\exp[i(\omega_k-\omega)\tau_0]\,-\,1}%
{i(\omega_k-\omega)\tau_0-0}\; e_{\mu}%
\label{3.4}%
\end{equation}%
describing the pulse propagating along the $z$-axis with carrier
frequency $\omega$, mode wave vector ${\mathbf k}\parallel z$,
mode frequency $\omega_k=c\,k$ and polarization ${\mathbf e}$. The
pulse wavefront arrives at the plane $z=0$ at the time $t=0$ and
$\alpha_{\mathbf k\mu}$ are the eigenstates of the mode
annihilation operator:
\begin{equation}%
a_{\mathbf k\mu}|\alpha_{\mathbf k\mu}\rangle = %
\alpha_{\mathbf k\mu}\,|\alpha_{\mathbf k\mu}\rangle%
\label{3.5}%
\end{equation}%
This initial coherent state gives an expected number of photons
arriving at the system during the interval $(0,\tau_0)$.

For an isotropic Gaussian-type cloud, with a radius normally on
the order of $1$ mm, the Green's Function (\ref{2.8}) can be
expressed in the form (\ref{A.6}), where for an isotropic medium
the slowly varying amplitude is given by
\begin{equation}%
X_{ij}(\mathbf{r}_{1},\mathbf{r}_{2},\omega)=%
\delta _{ij}^{\perp}\,%
\exp\left[-\frac{2\pi\omega}{c}\int_{{\mathbf r}_1}^{{\mathbf r}_2}\,%
\chi({\mathbf r},\omega)\,ds\right]%
\label{3.7}%
\end{equation}%
where the integral is evaluated along the ray $s$ linking the
points ${\mathbf r}_1$  and ${\mathbf r}_2$, and $\chi({\mathbf
r},\omega)$ is the local susceptibility of the inhomogeneous
medium, see definitions (\ref{A.13}), (\ref{A.14}).

In Figure \ref{Fig.1}(a) we show how the originally rectangular
pulse (\ref{3.4}) with duration of $\tau_0=20\,\gamma^{-1}$ is
distorted in the scattering response in the backward direction.
The calculations were made for the resonance $F_0=3\to F=4$
hyperfine transition of the $D_2$ line of ${}^{85}$Rb for the
optical thicknesses $b_0$ varying in the interval from $0$ to $5$.
During such a long probe of the sample the steady state behavior,
established within the temporal duration of the pulse, is clearly
indicated. But the sharp shapes of initial and final fronts of the
pulse are smoothed in the response pulse because of the delay in
multiple scattering of different orders. The dependence on optical
thickness emphasizes that this delay is longer as the scattering
in higher orders becomes more important, i. e. for higher $b_0$.
In turn, as shown in Figure \ref{Fig.1}(b), the presence of
different scattering orders in the outgoing response also
manifests themselves in multi-exponential decay of the
fluorescence, which approximately approaches the zero Holstein
mode for a rather long time and for high $b_0$.

The importance of the response inertia is remarkably indicated in
the time dynamics of the enhancement factor for the CBS process.
For a time dependent process the enhancement factor should be
defined by the ratio of the instantaneous intensity of light
scattered in the backward direction at the moment $t$ to the
contribution of the single scattering and only ladder-type terms
of the higher orders of multiple scattering
\begin{equation}%
\alpha(t)=1+\frac{I_C(t)}{I_S(t)+I_L(t)}%
\label{3.9}%
\end{equation}%
Here $I_S(t)$, $I_L(t)$, $I_C(t)$ are respectively the single,
ladder and crossed (interference) contributions to the
instantaneous outgoing intensity.

In Figure \ref{Fig.2} we illustrate the time dependence of the
enhancement factor calculated in the same conditions and for the
same parameters as the intensity response in Figure \ref{Fig.1}.
The most interesting is a break point on the graph corresponding
to the moment when the probe pulse is switched off. During the
decay process the enhancement factor rises up at the beginning
stage and drops down only after a delay and with a slower rate.
Such behavior is a direct consequence of the intensity decay shown
in Figure \ref{Fig.1}. Since the single scattering disappears
first then, in accordance with definition (\ref{3.9}), it should
be expected that the instantaneous value of the enhancement factor
will be raised immediately after the pulse is switched off.

\begin{figure}[tp]
{\includegraphics{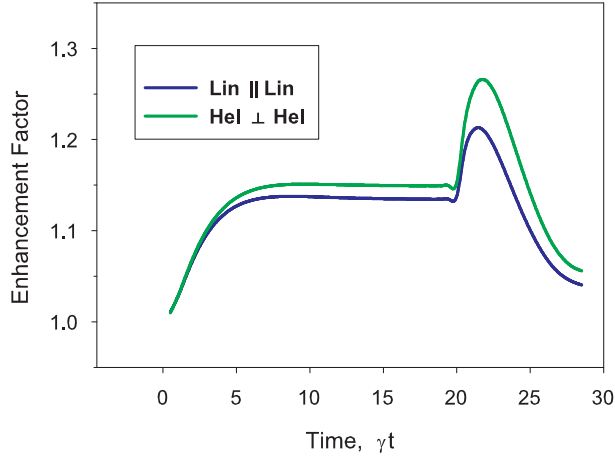}}
\caption{Time-dependence of the enhancement factor for the
excitation conditions of Figure \ref{Fig.1} in the lin $\|$ lin
parallel and hel $\bot$ hel scattering channels. The optical
thickness $b_0$ is equal to $5$ for this graph.}
\label{Fig.2}%
\end{figure}%

\begin{figure}[tp]
{\includegraphics{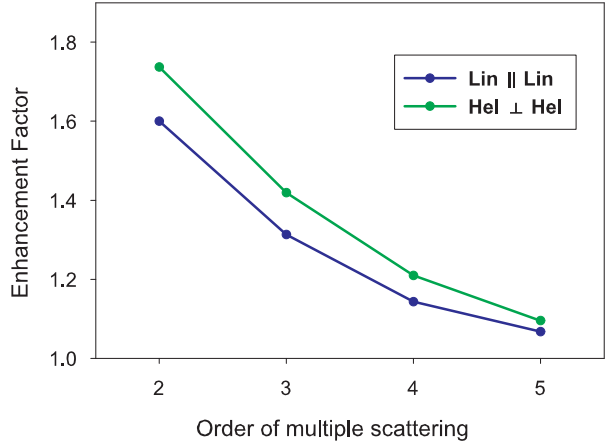}}
\caption{Partial contributions to the enhancement factor as
function of the order of multiple scattering for the same
polarization channels as in Figure \ref{Fig.2}.}
\label{Fig.3}%
\end{figure}

As one can see, the time dependent analysis gives us certain
spectroscopic access to the selective information about different
orders of multiple scattering. The CBS phenomenon could be an
important process for this. If the sensitivity of time dependent
measurements were high enough then the long-term decay of the
instantaneous value of the enhancement factor would strongly
select the contribution of different scattering orders. Roughly we
would say that the time decay shown in Figure \ref{Fig.2} just
copies the partial contribution to the enhancement coming from
different orders of multiple scattering. To show this in Figure
\ref{Fig.3} we plot the calculated data for the partial
contributions to the enhancement factor for the scattering orders
from two up to five. As one can see there is a qualitative
coincidence between the graphs showing the time decay and the
dependence on scattering order. Both the dependencies approach the
value of unity because in higher scattering orders the number of
non-interfering amplitudes enlarges faster than the number of
interfering ones.

In discussions of the CBS process in the literature
\cite{Sheng,LagTig} the cone profile itself is normally suggested
as an experimental reference dependence for selecting different
orders of multiple scattering. Such a spatial spectroscopy method
utilizes an idea that the higher orders contribute and are
localized near the peak point of the cusp-like CBS cone. As is
clear from the above discussion, in the time dependent
spectroscopy there is an alternative situation when higher orders
contribute in the wing of the time decay of the enhancement
factor. Thus the spatial spectroscopy and time dependent
spectroscopy could be complementary techniques for study of the
weak localization of light in disordered media.

\subsection{Measurements of time dependent diffuse light
intensity} To date, there have been no measurements of the time
dependence, in ultracold atomic gas samples, of coherently
backscattered light. However, there have been several experimental
studies, in ultracold gas samples, of the time dependence of
multiply scattered light \cite{Balik1,Fioretti,LabeyrieTime1}. In
this section we present some of our combined experimental and
theoretical results on the polarization dependence of multiple
light scattering in ultracold atomic $^{85}$Rb.

The essential experimental details \cite{Balik1} are discussed in
Section 3 of this report.  In these experiments, the intensity of
light emitted in a direction orthogonal to the excitation light
source is measured as a function of time and linear polarization
state. These dependences are illustrated in Figure 12, which shows
the measured intensities in two orthogonal polarization channels
for resonance excitation of the $F_0 = 3 \rightarrow F = 4$
hyperfine transition. In the figure, the scale of the peak
intensity in the lin $\|$ lin channel is about 10${^4}$ counts. In
this data, the transient build up is due to multiple scattering of
probe radiation after it has been switched on. The time scale for
the process can be seen more clearly in the expanded view in the
lower panel of Fig. 12. The solid curves in Fig. 12 represent
Monte-Carlo simulations of the scattering process. Other than the
overall intensity scale, there are no adjustable parameters in the
comparison, with the simulation input data consisting of the
measured trap density profile and peak optical depth.

\begin{figure}[tp]
{\includegraphics{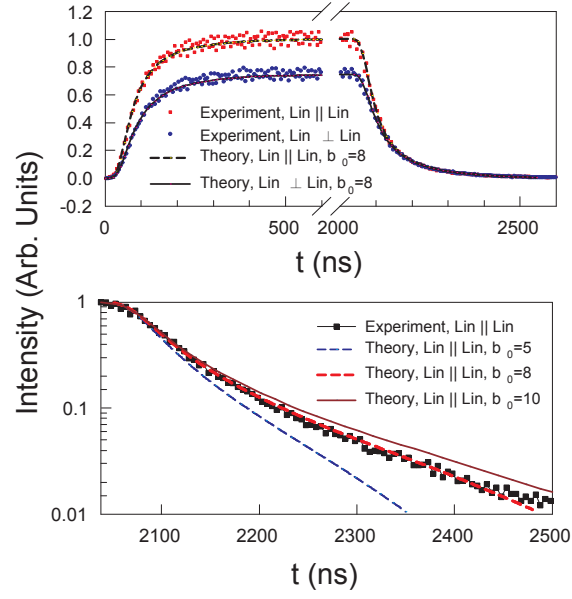}}
\caption{Measured time-dependent scattered light signal in
orthogonal linear polarization channels. Theoretical results are
indicated by the curves, as labelled in the figure legend.}
\label{Fig.7}%
\end{figure}

\begin{figure}[tp]
{\includegraphics{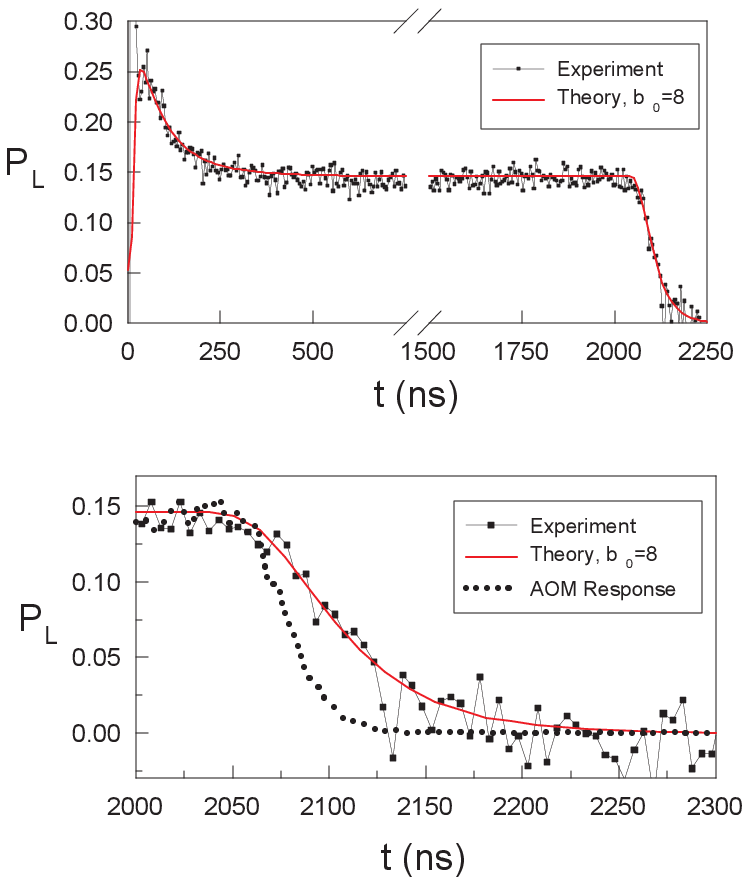}}
\caption{Measured time-dependent linear polarization degree, shown
as solid points. Monte Carlo simulation results are shown as solid
lines, while the limiting AOM response is shown as a dotted line.}
\label{Fig.8}%
\end{figure}%

This data is further reduced to emphasize the intensity
differences for the two orthogonal linear polarizations.  This is
done by defining a linear polarization degree as
\begin{equation}
P_{L} = \frac{I_{\parallel} - I_{\perp}}{I_{\parallel} +
I_{\perp}}%
\label{PL}%
\end{equation}
In the formula, $I_{\parallel}$ and $I_{\perp}$ represent the
measured intensities in the lin $\parallel$ lin and lin $\perp$
lin channels. The data in Fig. 12 then give the time-dependence of
$P_{L}$ shown in Fig. 13. There we see that $P_{L}$ enhances the
differences in the different channels, thus clearly showing the
time-dependent maximum in $P_{L}$ soon after the exciting pulse is
turned on, which is followed by approach to a steady state linear
polarization degree. We point out that the peak value of $P_{L}$
corresponds to a predominately single scattering value of $P_{L}$
= 0.268 for this resonance transition. As seen in the lower panel
of Fig. 13, once the probe laser is turned off, $P_{L}$ rapidly
decays toward 0. This emphasizes the quite different time scales
for decay of the population in comparison to that of the
electronic alignment.  In all of this data, the Monte-Carlo
simulations do an excellent job of describing both the light
intensity and polarization dynamics in the system.

Finally, it is important to note that the spectral dependence of
diffuse light scattering is strongly influenced by interferences
in the light scattering amplitudes from different excited
hyperfine levels.  This is illustrated in Figure 14, which shows
the spectral variation of the steady state linear polarization
degree.  In the figure, if hyperfine interferences were absent,
the single scattering atomic polarization would be very nearly
constant over the spectral range of the figure.  We should point
out that there is a much more gradual variation of the linear
polarization due to coherent excitation of the fine structure
multiplet levels.

\begin{figure}[tp]
{\includegraphics{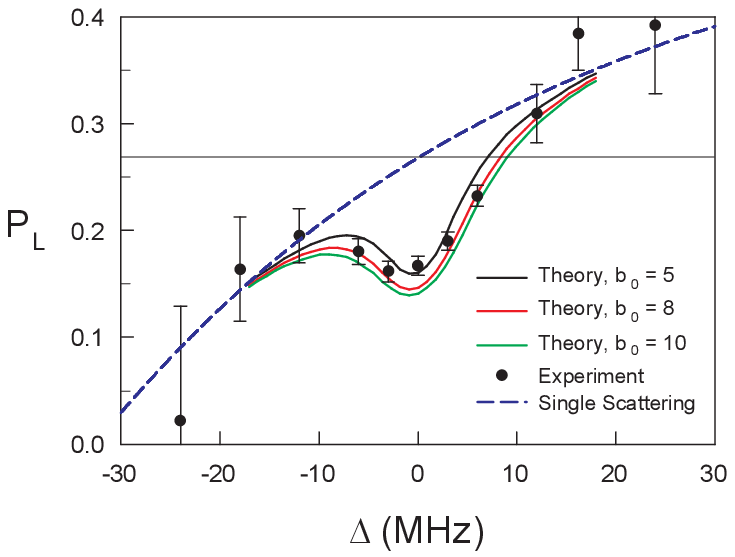}}
\caption{Detuning dependence of the linear polarization degree.
The horizontal line indicates the single scattering benchmark at
0.268. The dots represent the experimental data points, while the
blue chained curve is the expected variation for single atom
scattering. The red and blue curves indicate the polarization
variation for optical depths of $b_0 = 5$ (black), $b_0 = 8$
(red), and $b_0 = 10$ (green).}
\label{Fig.9}%
\end{figure}%

Instead, we see strong variations in the measured polarization due
to both multiple light scattering and due to hyperfine
interferences. For example, as the magnitude of the detuning is
made larger, the measured steady-state values for $P_{L}$ approach
the calculated frequency-dependent single scattering limit for
this transition. Near the resonance, the polarization is
significantly reduced by multiple coherent scattering. In all
cases, agreement with Monte Carlo simulations is very
satisfactory.

\subsection{Spectral distribution of the scattered light}

As an alternative to the time dependent analysis, it is possible
to determine the spectral selection of the multiple scattering
process. Different orders of the scattering should have different
spectral profiles in response to coherent monochromatic
excitation. In an experiment, precise selection can be done by
using the light beating spectroscopy technique \cite{Light
beating}. In this approach, mixing the scattered light with a
local oscillator wave in a heterodyne detection scheme yields a
photocurrent spectrum which will reproduce the spectrum of the
scattered light given by (\ref{3.2}) and (\ref{3.3}).

This spectrum can be calculated analytically if the following
three basic criteria are met. (i) The velocity distribution for
the atoms in the trap should be of the Maxwell-Boltzmann type.
(ii) Near the Doppler cooling limit the corresponding Doppler
frequency shift is much less than $\gamma$ and the retarded and
advanced type Green functions should be insensitive to the atomic
velocity distribution. (iii) For the same reason, the dependence
of the scattering amplitude on atomic velocity should be also
negligible for deeply cooled atoms. Then for the $N$-th order of
multiple scattering the partial spectral profile will be centered
at the either the Rayleigh elastic or Raman shifted outgoing
frequency $\omega_R$ and will be given by
\begin{equation}%
I_{N}(\omega)=\left\langle I_{N}\,%
\sqrt{\frac{2\pi}{\Gamma_N}}\,%
\exp\left[-\frac{(\omega-\omega_{R})^2}{2\Gamma_N^2}\right]\right\rangle%
\label{3.10}%
\end{equation}%
where the Gaussian bandwidth is given by the following sum
\begin{equation}%
\Gamma_N=\frac{1}{\sqrt{2}}\left[\sum_{i=1}^{N}|\Delta{\mathbf k}_i|^2\right]^{1/2}{v_0}%
\label{3.11}%
\end{equation}%
Here $\Delta{\mathbf k}_i$ with $i=1\div N$ is the sequence of
changes of the wave vector for the scattering of the light wave
along any randomly selected scattering chain consisting of $N$
atoms. Here the velocity $v_0$ is the most probable velocity for
the respective Maxwell-Boltzmann distribution, see definition in
section \ref{S4.2}. $I_N=I_N({\mathbf r}_1,\ldots{\mathbf r}_N)$
is the total intensity of the fraction of the light scattering in
the direction of observation by the subsequent scattering from
atoms located at spatial points ${\mathbf r}_1,\ldots{\mathbf
r}_N$. The angle brackets in (\ref{3.10}) denote the averaging
over the spatial distribution of these atoms. Such an averaging
extends over $\Delta{\mathbf k}_i$ and $\Gamma_N$ as well, which
also depend on the locations of atoms, see definitions
(\ref{2.3}), (\ref{2.4}). Then the total spectrum of the scattered
light is given by
\begin{equation}%
I(\omega)=\sum_N I_N(\omega)%
\label{3.12}%
\end{equation}%
It is remarkable that this result is valid for the scattering in
any direction including backscattering. The interference terms
have the same spectral profile as the ladder terms if the atoms do
not noticeably change their location on the spatial scale of
$\lambda/2\pi$ during the retardation delay while the light wave
passes through the scattering chain. As one can see in higher
orders of multiple scattering the spectrum becomes broader and for
an isotropic sample the bandwidth $\Gamma_N$ is enhanced by $N$.

However in reality the assumptions (i) - (iii) are not exactly
fulfilled. First, the velocity distribution is only approximately
Maxwell-Boltzmann-type because of complexity of the cooling
mechanism in MOT. Second, if the Doppler scale $kv_0$ is
comparable with the natural line width $\gamma$, there will be
additional damping mechanisms acting via the Green's propagator
phase, reducing the scattering amplitude. Thus the sum of Gaussian
profiles (\ref{3.12}) gives us only a convenient zero-level
approximation, or a certain type of reference dependence for
further spectroscopic analysis. The deviations from this basic
dependence provide us with actual information about real atomic
motion and spatial correlations of atoms in the sample.

\begin{figure}[tp]
{\includegraphics{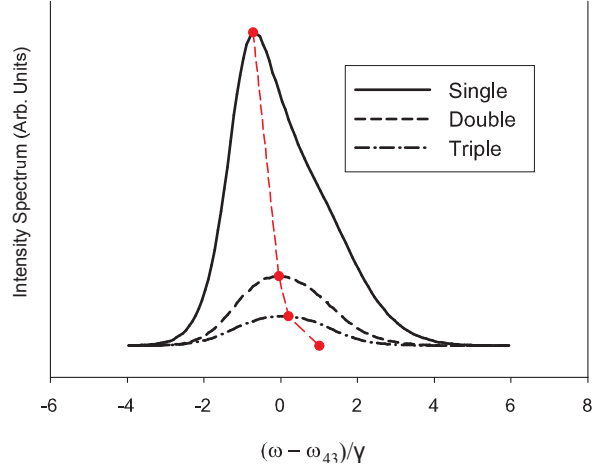}}
\caption{Spectral profiles of the light backscattered on the
$F_0=3\to F=4$ hyperfine transition of ${}^{85}$Rb in
Rayleigh-type lin $\parallel$ lin polarization channel for
different orders of multiple scattering. The probe light frequency
is shifted to the blue wing of the resonance by one natural line
width $\gamma$ and the velocity distribution of atoms is assumed
to be Maxwell-Boltzmann-type with $kv_0=\gamma$. The red circle
indicate how the maxima approach the limit of elastic scattering
$\omega_R=\omega_{43}+\gamma$ in higher orders.}
\label{Fig.4}%
\end{figure}%

We illustrate this in Figure \ref{Fig.4} where we show typical
dependencies of the spectral response on the monochromatic probe
wave for the scattering near the backward direction in the lin
$\parallel$ lin polarization channel from atoms of ${}^{85}$Rb.
The data are presented for the $F_0=3\to F=4$ hyperfine transition
of the $D_2$-line and for the different orders of multiple
scattering. The frequency of the probe wave is tuned to the blue
wing from the atomic resonance by one natural line width
$\omega_{43}+\gamma$. The velocity distribution is assumed to be
Maxwellian with $kv_0=\gamma$. For such "hot" atoms the
ladder-type response is mainly important in the detection channel.
For Rayleigh-type scattering initiated on the selected closed
transition the carrier frequency is $\omega_R=\omega_{43}+\gamma$
and, according to (\ref{3.10}), it is expected that the spectral
response should be centered and distributed near this carrier
frequency shifted to the {\it blue wing} of resonance. But as one
can see this is not the case and the mean frequency of the
scattered light is shifted to the {\it red wing} of the atomic
resonance. This is a typical manifestation of the Doppler effect
in the denominators of scattering amplitudes. In a single
scattering event, the scattering is preferably organized from the
atoms moving with $v_z\sim \gamma/k$ in the direction of the
incoming wave front. Thus the outgoing wave, which is scattered
from these atoms, will be shifted by $-2kv_z\sim -2\gamma$. In a
weaker form such an effect is preserved for the double and triple
scattering channels, as also shown by the corresponding
dependencies of Figure \ref{Fig.4}.

\begin{figure}[tp]
{\includegraphics{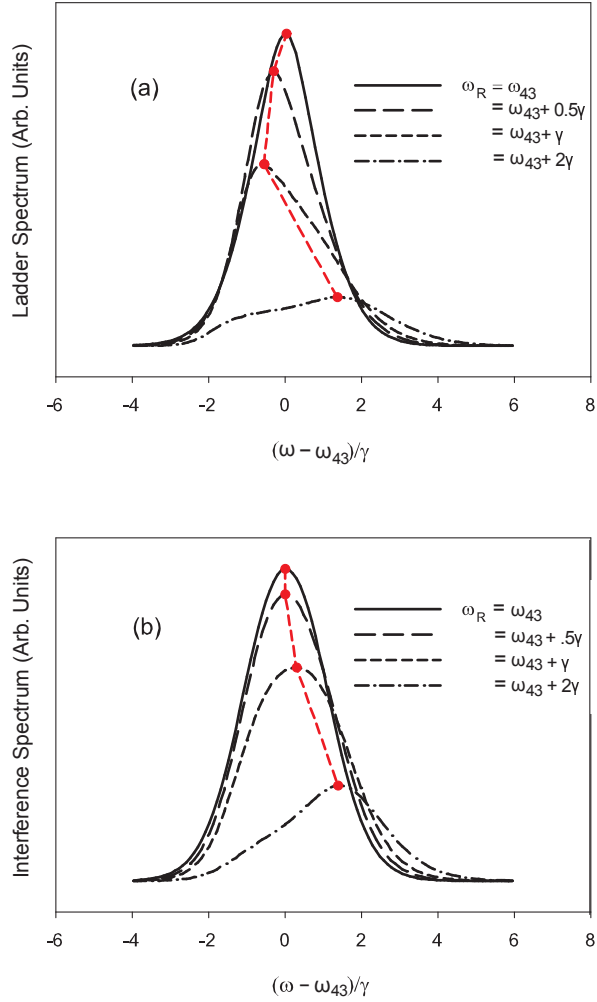}}
\caption{Spectra of the ladder (a) and interference (b)
contributions to the light backscattered on the $F_0=3\to F=4$
hyperfine transition of ${}^{85}$Rb for different frequency
offsets of the probe light from the resonance transition. Other
parameters are the same as in Figure \ref{Fig.4}. The red circles
trace the location of the maxima as they approach the elastic
carrier frequency for far off-resonant scattering.}
\label{Fig.5}%
\end{figure}

In Figure \ref{Fig.5} we show how the spectral profiles for the
total ladder and interference contributions depend on the
frequency of the probe light, which is varying from the resonance
transition frequency $\omega_{43}$ up to $\omega_{43}+2\gamma$.
Other parameters are the same as in Figure \ref{Fig.4}. As is
clearly seen, the location of the maximum for the ladder portion
shifts to the red wing for small detuning and approaches the
frequency of elastic scattering $\omega_R$ only for large
detunings. For interference terms, the location of the maximum
shifts approximately as $\omega_R$, which is because of no single
order contribution in this case.

If atoms are cooled close to the Doppler limit, such that $kv_0\ll
\gamma$, the spectral profile will be given by expressions
(\ref{3.10})-(\ref{3.12}). In this case its spectral bandwidth can
be quite narrow and in an experiment a high quality monochromatic
local oscillator wave should be applied to resolve the spectrum by
heterodyne detection. But the possible experimental realization
could give us certain access to observation of the actual velocity
distribution of the ultracold atoms in the trap, which is not
necessarily Maxwellian.

\subsection{Polarization-sensitive effects: Anti-enhancement in
weak localization regime}

Let us consider an atomic ensemble consisting of atoms oriented in
their angular momenta. Further, let this ensemble be probed with
circular polarized light in a direction antiparallel to the
magnetization direction of the atomic vapor. Such a geometry
requires special preparation since, because of the optical pumping
process, there is a tendency to reorient the collective spin
vector of the atomic ensemble along the beam, especially following
a long interaction with the probe light, and after an accumulation
of a sufficient number of Raman transitions. However during the
short pulsed excitation we can neglect the optical pumping
mechanism and assume that most of the atoms populate the
$|F_0,m=-F_0\rangle$ Zeeman state. In addition, and for the reason
explained below, we can assume that the ensemble is located in an
external weak magnetic field directed along the light beam. Thus
the photons scattered via Raman channels will be generally Raman
shifted. This splitting can be made quite small and less than the
natural line width of the respective optical transition but still
resolved with high resolution spectroscopy techniques.

\begin{figure}[tp]
{\includegraphics{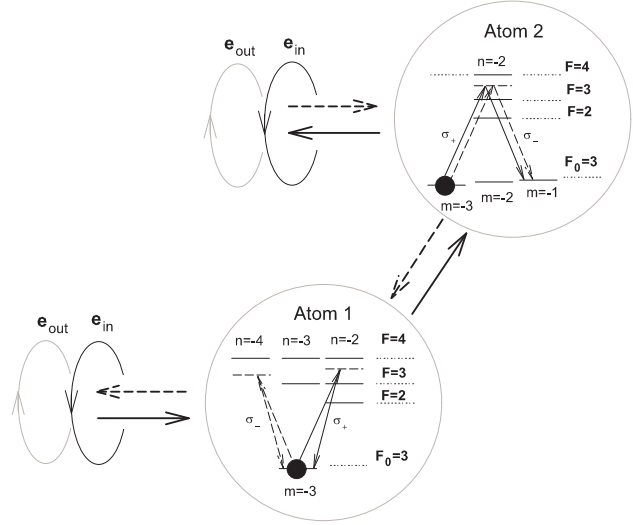}}
\caption{Diagram explaining the anti-localization phenomenon in
the helicity preserving scattering channel for double scattering
of circular polarized light from an ensemble of ${}^{85}$Rb atoms
oriented opposite to the helicity vector of the probe beam. The
destructively interfering amplitudes are a combination of
Rayleigh- and Raman-type transitions. The solid and dashed lines
indicate the interfering direct and reciprocal scattering paths
for probing between $F_0 = 3 \to F = 4$ and $F_0 = 3 \to F = 3$
hyperfine transitions.}
\label{Fig.6}%
\end{figure}

In Figure \ref{Fig.6} we show the double backscattering of
incident light of positive helicity on a system consisting of two
${}^{85}$Rb atoms; the exit channel consists of detection of light
also of positive helicity. The two interfering channels, which are
shown here, repopulate atoms via Raman transitions from the
$F_0,m=-F_0$ to the $F_0,m=-F_0+2$ Zeeman sublevel. In the direct
path the scattering consists of a sequence of Rayleigh-type
scattering in the first step and of Raman-type scattering in the
second. In the reciprocal path, Raman-type scattering occurs
first, and the positive helicity photon undergoes Rayleigh-type
scattering in the second step. Since identical helicities in
incoming and outgoing channels have opposite polarizations with
respect to the laboratory frame, there is an important difference
in transition amplitudes associated with the Rayleigh process for
these two interfering channels. Indeed, in the direct path the
$\sigma_{+}$ mode is coupled with $F_0 = 3 \to F = 4$, $F_0 = 3
\to F = 3$ and $F_0 = 3 \to F = 2$ hyperfine transitions. But in
the reciprocal path, the $\sigma_{-}$ mode can be coupled only
with the $F_0 = 3 \to F = 4$. As we see from the diagrams shown in
Figure \ref{Fig.6}, where the probe light frequency is scanned,
for example, between the $F_0 = 3 \to F = 4$ and $F_0 = 3 \to F =
3$ transitions, a unique spectral feature is found when the
scattering amplitudes connecting the direct and reciprocal
scattering channels are equal in absolute value but have phases
shifted by an angle close to $\pi$. From an electrodynamic point
of view, such conditions are realized when, due to the asymmetry
in the Rayleigh-type transitions, the real part of the
susceptibility of the sample is positive for the $\sigma_{-}$ mode
and is negative for the $\sigma_{+}$ mode. Since, in a first
approximation, the amplitudes of the processes shown in Figure
\ref{Fig.6} have opposite signs they will interfere destructively
with anti-enhancement of the light scattered in the backward
direction in the helicity preserving channel. Such an unusual
behavior in the CBS process is connected with the Raman nature of
the helicity preserving scattering channel. To observe this
effect, and not have it obscured by other competitive and
constructively interfering channels, both special light
polarization and selection of certain atomic $\Lambda$-type
transition are required.

In our example, and in the case of degenerate Zeeman sublevels,
there are several competing channels of double scattering, which
can interfere constructively. In an experiment, possible selection
of either destructively or constructively interfering channels can
be done by focusing attention to their angular dependence with
respect to the location of the atomic scatterers. For an optically
thin sample the process of Figure \ref{Fig.6} dominates for atoms
located preferably along the probe beam direction. The respective
angular factor is proportional to the probability to initiate
$\sigma_{+}$ or $\sigma_{-}$ transitions on the second atoms by
the photon also emitted on the $\sigma_{+}$ or $\sigma_{-}$ by the
first atom. The probability is given by
\begin{equation}
P_{++}(\theta)=P_{--}(\theta)\propto%
\frac{1}{4}(\cos^2\theta\,+\,1)^2%
\label{3.13}%
\end{equation}%
In turn, other processes are dominant if atoms are located
preferably in the plane orthogonal to the probe beam. Thus in
double scattering channel one can expect that, for a cigar-type
atomic cloud, stretched along the probe beam direction and
squeezed in orthogonal directions, the destructively interfering
channels should give the dominant contribution.

In Figure \ref{Fig.7} we show the dependence of the enhancement
factor, calculated only for double scattering amplitudes, on the
frequency of the probe field $\omega_L$. The blue curve is the
contribution of the amplitudes expressed by the diagrams of Figure
\ref{Fig.6}. The green curve includes the whole set of the double
scattering amplitudes including all the constructively interfering
channels. The calculations were made for the optical depth near
the threshold level of $b_{0} \sim 1$ at each spectral detuning
$\Delta=\omega_L-\omega_{43}$. It may seem surprising that even in
this hypothetical situation, where only the process of Figure
\ref{Fig.6} contributes to the enhancement factor, it actually
never drops down to zero level. This is because of the complexity
of the Green propagation function in the spin polarized gas, which
is discussed in an appendix. If locations of the atoms are shifted
in the transverse plane, the phases of the $\sigma_{+}$ and
$\sigma_{-}$ modes in the intermediate segments of the direct and
reciprocal paths will be mismatched because of the refraction
anisotropy of the sample. It is also unexpected that the
accumulation of all the constructively interfering channels do not
overwhelm the anti-localization effects in the observation
conditions. The enhancement factor stays less than unity in the
level of twenty percent in the total double scattering outcome. As
follows from Figure \ref{Fig.7} this takes place near the
$F_0=3\to F=2$ resonance.

\begin{figure}[tp]
{\includegraphics{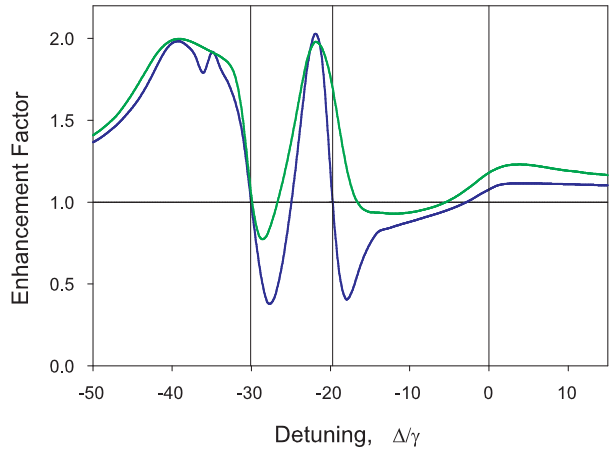}}
\caption{The enhancement factor in the helicity preserving channel
for double scattering as a function of detuning
$\Delta=\omega_L-\omega_{43}$ for the probe laser scanning the
upper hyperfine manifold of ${}^{85}$Rb. The blue curve shows the
contribution of the processes depicted in Figure \ref{Fig.6} and
in the green curve all the constructively interfering channels
were added. Vertical lines indicate the location of hyperfine
resonances.}
\label{Fig.7}%
\end{figure}%

Let us briefly discuss the feasibility of how the
anti-localization phenomenon could be observed in experiment. Our
previous proposal \cite{KSH} was based on the idea to organize an
inelastic Raman type selection of the process shown in Figure
\ref{Fig.6}. For this the Zeeman sublevels of the ground state
could be split by external magnetic and electric fields as shown
in the energy diagrams of Figure \ref{Fig.6}. Then the heterodyne
detection seems a quite preferable high resolution spectroscopic
technique to organize such a selection. Here we draw attention to
another possibility to preferably select the double scattering,
this by the time dependent spectroscopy technique described in
Section \ref{3A}. As far as the decay rate of an atomic dipole is
insensitive to its excitation frequency, the different orders of
the multiple scattering will be subsequently observed during the
time decay after the probe radiation is switched off, see Figure
\ref{Fig.2}. This means that in the anti-localization regime the
time dependence of the enhancement factor should be expected to
break down (not up) after the probe pulse switched off. Our
evaluations reproduced by the graphs of Figure \ref{Fig.7} show
that such a challenging and interesting scattering phenomenon as
anti-localization of light is readily within current experimental
capability.

\section{Conclusion}
In this review, we have provided a detailed summary of theoretical
and experimental developments associated with coherent
backscattering of light from ultracold samples of Rb or Sr atoms.
The main observables are the coherent backscattering enhancement
and the width of the backscattering cone. The paper considers the
influence on these observables of a number of external parameters,
including spectral variation from resonance excitation, the
influence of magnetic fields, suppression of the coherent
backscattering effect due to stronger electromagnetic fields. In
the spectral variations, interference effects due to normally
energetically distant hyperfine transitions are considered.
Interference effects due to coherent scattering from nondegenerate
levels are considered, including a regime of so-called
antilocalization.   The time evolution of the coherently
backscattered light is also discussed, including the influence on
the coherent backscattering cone with time.  The method of light
beating spectroscopy is discussed in the context of its utility in
extracting additional information, including the actual velocity
distribution of the atoms, regarding the light scattering process.

We emphasize that research exploring the physics of weak
localization in ultracold atomic gases is clearly in the earliest
stages, and there remain many unexplored lines of inquiry.  Among
these are fundamental aspects of nonlinear optical phenomena in
the multiple scattering regime, and investigation of multifield
coherences such as electromagnetically induced transparency and
absorption. Research into nonlinear optics at very low light
levels and the possibilities associated with generation of novel
types of polaritonic excitations may have important practical
applications in the area of quantum information processing.
Finally, the regime of strong localization of light in ultracold
gases is within reach of current experimental techniques of
ultracold atomic physics.  Strong light localization, as a phase
transition in the quantum properties of energy transport in
ultracold gases, promises fascinating opportunities for new
experimental and theoretical insights into the physics of
disordered systems.

\section*{Acknowledgments}
We would like to thank Prof. Sergey Kulik whose kindly suggested
to us to prepare this review. We appreciate the financial support
of the National Science Foundation (NSF-PHY-0355024), the Russian
Foundation for Basic Research (RFBR-05-02-16172-a), and the North
Atlantic Treaty Organization (PST-CLG-978468). D.V.K. would like
to acknowledge the financial support from the Delzell Foundation,
Inc.

\appendix\section{Green's propagation function}
\setcounter{equation}{0}%

\subsection{Dyson equation}
Light propagation through a sample consisting of atoms with
arbitrary polarizations of their angular momenta was considered
first by Cohen-Tannoudji and Lalo\"{e} in Refs. \cite{ChTnLa}. In
those papers, such well known effects of optical anisotropy of an
atomic vapor as birefringence, gyrotropy and dichroism were linked
with the formalism of the irreducible tensor components of atomic
polarization, i.e. with the orientation vector and alignment
tensor. In the formalism of the Green's function, the problem of
light propagation through a polarized atomic vapor was discussed
in Ref. \cite{KSS}. The technique developed there permitted
analytical solution for the Green's function in many practically
important applications. We review here the basic results of
\cite{KSS} and make them closer to the previously discussed
conditions of CBS experiments with a cold atomic vapor confined to
a MOT.

According to a general quasi-particle conception, the
retarded-type Green's function can be found as a solution of the
following Dyson equation
\begin{eqnarray}
&&\left[\frac{1}{c^2}\frac{\partial^2}{\partial\tau^2}-\triangle_1\right]%
D_{ij}^{(R)}(\mathbf{r}_{1},\mathbf{r}_{2},\tau)=%
-4\pi\hbar\delta(\tau)\delta^{\bot}_{ij}(\mathbf{r}_{1}-\mathbf{r}_{2})%
\nonumber\\%
\lefteqn{+\int_0^\infty\!d\tau'\int\! d^3r_1'%
({\cal P}_{\bot}^{(R)})_{ii'}(\mathbf{r}_{1},\mathbf{r}_{1}',\tau')\,%
D_{i'j}^{(R)}(\mathbf{r}_{1}',\mathbf{r}_{2},\tau-\tau')}%
\nonumber\\%
&&%
\label{A.1}%
\end{eqnarray}%
which reveals the normal vacuum wave equations modified by the
contribution of the polarization operators on the right-hand side.
Rigorously speaking, this equation defines only the positive
frequency components of the Green's function derived under the
assumptions of the rotating wave approximation. But such an
assumption is consistent with the general restrictions of our
calculations and is just the Fourier image of the positive
frequency components of the Green's function, which has to be
substituted into Eq.(\ref{2.8}).

An important characteristic of equation (\ref{A.1}) is the
transverse delta-function, which is given by
\begin{equation}
\delta^{\bot}_{ij}({\mathbf r})= \delta_{ij}\,%
\delta({\mathbf r})\,%
+\,\frac{1}{4\pi} \frac{\partial^2}{\partial x_i\partial x_j}\,%
\left(\frac{1}{r}\right)%
\label{A.2}%
\end{equation}%
Its Fourier components coincide with the
$\delta^{\bot}_{ij}$-symbol defined by Eq.(\ref{2.6}). This
function selects, from any tensor characteristic connected with
atomic variables, the transverse projection. Such a projected
kernel of the polarization operator enters the Dyson equation
(\ref{A.1}) and is given by
\begin{eqnarray}
({\cal P}_\bot^{(R)})_{ii'}({\mathbf r},{\mathbf r}';\tau)&=&%
\int\!\int d^3r_1d^3r_1'%
\delta^{\bot}_{ij}({\mathbf r}-{\mathbf r}_1)%
\nonumber\\%
&&\times {\cal P}^{(R)}_{jj'} ({\mathbf r}_1 ,{\mathbf r}_1' ;\tau)\,%
\delta^{\bot}_{j'i'}({\mathbf r}_1'-{\mathbf r}')%
\nonumber\\%
&&\label{A.3}%
\end{eqnarray}%
In steady state conditions the polarization operator can be
expressed by its Fourier transform
\begin{equation}
{\cal P}^{(R)}_{ii'}({\mathbf r},{\mathbf r}';\tau)=%
\int\!\frac{d\omega}{2\pi}\,e^{-{\rm i}\omega\tau}\,%
{\cal P}^{(R)}_{ii'}({\mathbf r},{\mathbf r}';\omega)%
\label{A.4}%
\end{equation}%
where
\begin{eqnarray}
\lefteqn{{\cal P}^{(R)}_{ii'}({\mathbf r},{\mathbf r}';\omega)=%
-\frac{4\pi\omega^2}{c^2} \sum_{n}\sum_{m,m'}%
(d_{i})_{mn}(d_{i'})_{nm'}}%
\nonumber\\%
 &&\times\int\!\int\!
\frac{d^3p}{{(2\pi\hbar)}^3} \frac{d^3p'}{{(2\pi\hbar)}^3}%
\exp[-\frac{i}{\hbar}({\mathbf p}-{\mathbf p'})({\mathbf r}-{\mathbf r}')]%
\nonumber\\%
&&\times\frac{1}{\hbar(\omega-\omega_{nm})\,-\,%
\epsilon(p')+\epsilon(p)\,+\,{\rm i}\hbar\gamma_{n}/2}%
\nonumber\\%
&&\times\rho_{m'm}\left(\frac{{\mathbf p}+{\mathbf
p}'}{2},\frac{{\mathbf r}+{\mathbf r}'}{2}\right)%
\label{A.5}%
\end{eqnarray}%
Here $\rho_{m'm}({\mathbf p},{\mathbf r})$ are the steady state
components of atomic density matrix, in the Wigner representation,
generated by an optical pumping process or due to other physical
mechanisms initiating the spin polarization. We will further
assume any possible polarization in degenerate or quasi-degenerate
system of Zeeman sublevels. The function $\epsilon(p)$ denotes the
remaining atomic kinetic energy.

In application to the CBS process we only need to know the long
range asymptotic form of the Fourier components of the retarded
Green's function, see relation (\ref{2.8}), where the spatial
points ${\mathbf r}_1$ and ${\mathbf r}_2$ are separated by a
distance of many wavelengths. Then the Green's function can be
factorized into the following product of the rapidly oscillating
exponential and the slowly varying amplitude $X_{ij}({\mathbf
r}_1,{\mathbf r}_2;\omega)$
\begin{equation}%
D_{ij}^{(R)}(\mathbf{r}_{1},\mathbf{r}_{2},\omega)=%
-\hbar X_{ij}({\mathbf r}_1,{\mathbf r}_2;\omega)\,%
\frac{\exp[ik|{\mathbf r}_1-{\mathbf r}_2|]}%
{|{\mathbf r}_1-{\mathbf r}_2|}%
\label{A.6}%
\end{equation}%
where $k=\omega/c$. The slowly varying amplitudes satisfy the
following differential equations
\begin{eqnarray}
\frac{\partial}{\partial z_1} X_{ij}({\mathbf r}_1 ,{\mathbf r}_2 ;\omega)&=&%
\frac{2\pi{\rm i}\omega}{c}\sum_{i'={x,y}}%
\chi_{ii'}({\mathbf r}_1,\omega)\,%
X_{i'j}({\mathbf r}_1 ,{\mathbf r}_2 ;\omega)%
\nonumber\\%
X_{ij}({\bf r}_1 ,{\bf r}_2 ;\omega)&\to& \delta^{\bot}_{ij}\ \ {\rm at}\ \ %
z_1\to z_2%
\label{A.7}%
\end{eqnarray}%
where $\chi_{ii'}({\mathbf r},\omega)$ is the tensor of the local
dielectric susceptibility of the inhomogeneous and anisotropic
medium, which is given by
\begin{eqnarray}
\chi_{ii'}({\mathbf r},\omega)&=&-\sum_{n}\sum_{m,m'}%
(d_{i})_{mn}(d_{i'})_{nm'}%
\nonumber\\%
 &&\times\int\!\frac{d^3p}{{(2\pi\hbar)}^3}\,%
\frac{\rho_{m'm}({\mathbf p},{\mathbf r})}%
{\hbar(\omega-\omega_{nm}-{\bf k}{\bf p}/\mu)\,+\,%
{\rm i}\hbar\gamma_n/2}%
\nonumber\\%
&&\label{A.8}%
\end{eqnarray}%
where $\mu$ is the atomic mass. The sum over tensor indices in
equation (\ref{A.7}) is extended only over transverse components
$x,y$ in the reference frame associated with the $z$-direction
along the ray linking the points ${\mathbf r}_1$ and ${\mathbf
r}_2$. The indices $i,j$ can be also equal to $x,y$ in this frame.

\subsection{Representation of irreducible tensor components}
The basic equation (\ref{A.7}) can be modified to a form more
convenient for further analysis by expanding the atomic density
matrix in terms of irreducible tensor components. As a first step
we factorize the Wigner density matrix into the product
\begin{equation}
\rho_{m'm}({\mathbf p},{\mathbf r})=%
\tilde{\rho}_{m'm}({\mathbf r})\,%
f({\mathbf p},{\mathbf r})%
\label{A.9}%
\end{equation}%
where $f({\mathbf p},{\mathbf r})$ is the classical distribution
function in phase space and $\tilde{\rho}_{m'm}({\mathbf r})$ are
the density matrix elements of the internal states.

>From here let us restrict our discussion by the practically
important assumption that the spatial dependence of the atomic
polarization $\tilde{\rho}_{m'm}({\mathbf r})$ has only a
negligible change along the spatial scale associated with an
average photon free path in the sample. Then, as a good
approximation in solving equations (\ref{A.7}), we can neglect the
dependence on ${\mathbf r}$ and consider the $\tilde{\rho}_{m'm}$
to be constant parameters. Then instead of a Zeeman basis, one can
introduce the representation of irreducible tensor components by
the following expansion
\begin{equation}
\tilde{\rho}_{kq}^{F_0}=\sqrt{\frac{2k+1}{2F_0+1}} \sum_{m,m'}%
C_{F_0m'\,kq}^{F_0m}\, \tilde{\rho}_{m'm}%
\label{A.10}%
\end{equation}%
where $F_0$ is the total angular momentum of the ground state and
$C_{\ldots\,\ldots}^{\ldots}$ is a Clebsch-Gordon coefficient in
the notation of Ref. \cite{VMK}.

Let us define the basis set of circular polarizations. The co- and
contravariant components of any complex vector ${\mathbf
\varepsilon}$ in the plane orthogonal to the $z$-direction are
given by
\begin{eqnarray}
\varepsilon_{+1}&=&-\varepsilon^{-1}\;=\;-\frac{1}{\sqrt{2}}%
(\varepsilon_x\,+\,{\rm i}\varepsilon_y)%
\nonumber\\
\varepsilon_{-1}&=&-\varepsilon^{+1}\;=\;+\frac{1}{\sqrt{2}}%
(\varepsilon_x\,-\,{\rm i}\varepsilon_y)%
\label{A.11}%
\end{eqnarray}%
In the basis set of circular polarizations the susceptibility
tensor projected on the $x,y$- plane can be expanded in the set of
identity and Pauli matrices as
\begin{equation}
\chi_{q}{}^{q'}({\mathbf r},\omega)=%
\chi({\mathbf r},\omega)\left[\rho_0\delta_{q}{}^{q'}\,+\,%
(\mbox{\boldmath $\rho$}\,%
\hat{\mbox{\boldmath $\sigma$}})_{q}{}^{q'}\right]%
\label{A.12}%
\end{equation}%
where the symbolic vector $\hat{\mbox{\boldmath
$\sigma$}}=\hat{\sigma}_x,\hat{\sigma}_y,\hat{\sigma}_z$ is the
set of Pauli matrices. In this expansion the upper row and left
column of Pauli matrices are associated with the $+1$ index and
the respective lower row and right column with the $-1$ index.

The local susceptibility of an isotropic medium $\chi({\mathbf
r},\omega)$ can be expanded in a sum of partial contributions for
each $F_0\to F$ transition
\begin{equation}%
\chi({\mathbf r},\omega)=\sum_{F}\,\chi_{F_0F}({\mathbf r},\omega)%
\label{A.13}%
\end{equation}%
where the partial contribution is given by
\begin{eqnarray}
\chi_{F_0F}({\mathbf r},\omega)&=&-\frac{|d_{F_0F}|^2}{3(2F_0+1)}%
\int\!\frac{d^3p}{{(2\pi\hbar)}^3}%
\phantom{f({\mathbf p},{\mathbf r})\ (3.14)}
\nonumber\\%
&&\times\frac{f({\mathbf p},{\mathbf r})}%
{\hbar(\omega-\omega_{FF_0}-{\mathbf k}{\mathbf v})+{\rm i}\hbar\gamma_{F}/2}%
\label{A.14}%
\end{eqnarray}%
where $d_{F_0F}$ are the reduced matrix elements of the dipole
moment for $F_0\to F$ transition.

Other expansion parameters $\rho_0$ and $\mbox{\boldmath
$\rho$}=(\rho_x,\rho_y,\rho_z)$ introduced by expansion
(\ref{A.12}) are subsequently given by
\begin{eqnarray}
\rho_0&=&\rho_0(\omega)\;=\;1\,+\,c_2(\omega)\,\frac{1}{\sqrt{6}}\,%
\tilde{\rho}_{20}^{F_0}%
\nonumber\\%
\rho_x&=&\rho_x(\omega)\;=\;c_2(\omega)\,%
\frac{1}{2}\left[\tilde{\rho}_{2-2}^{F_0}+\tilde{\rho}_{22}^{F_0}\right]%
\nonumber\\%
\rho_y&=&\rho_y(\omega)\;=\;c_2(\omega)\,%
\frac{1}{2{\rm i}}\left[\tilde{\rho}_{2-2}^{F_0}-\tilde{\rho}_{22}^{F_0}\right]%
\nonumber\\%
\rho_z&=&\rho_z(\omega)\;=\;c_1(\omega)\,\tilde{\rho}_{10}^{F_0}%
\label{A.15}%
\end{eqnarray}%
where the complex factors $c_1(\omega)$ and $c_2(\omega)$ are
\begin{eqnarray}
c_1(\omega)&=&\frac{1}{\chi({\mathbf r},\omega)}\sum_{F}%
(-)^{F+F_0}\frac{3}{\sqrt{2}}(2F_0+1)%
\nonumber\\%
&&\times\left\{\begin{array}{ccc} 1&1&1\\%
F_0&F_0&F\end{array}\right\}%
\chi_{F_0F}({\mathbf r},\omega)%
\nonumber\\%
&&\nonumber\\%
c_2(\omega)&=&\frac{1}{\chi({\mathbf r},\omega)}\sum_{F}%
(-)^{F+F_0+1}3(2F_0+1)%
\nonumber\\%
&&\times\left\{\begin{array}{ccc} 1&1&2\\%
F_0&F_0&F\end{array}\right\}%
\chi_{F_0F}({\mathbf r},\omega)%
\label{A.16}%
\end{eqnarray}%
These generally complex parameters do not depend on ${\mathbf r}$
if the classical distribution $f({\mathbf p},{\mathbf r})$ is
factorized into a product of independent spatial and velocity
distributions, and they become real in the case of a closed
$F_0\to F$ transition. In experiments carried out with ultracold
atoms, and with high spectral selection of certain hyperfine
transitions, these factors have only a small admixture of
imaginary part. For an isotropic sample, $\rho_0\to 1$ and
$\mbox{\boldmath $\rho$}\to 0$.

\subsection{The phase integral representation of slowly varying
amplitudes}%
The independence of the symbolic vector $\mbox{\boldmath $\rho$}$
on ${\mathbf r}$ leads to commutativity of matrices
$\hat{\chi}({\mathbf r},\omega)$ considered at different spatial
points along a light ray. In turn, this makes possible analytical
solution of equation (\ref{3.7}). Let us define the complex length
$\rho$ of the symbolic vector $\mbox{\boldmath $\rho$}$ as
\begin{eqnarray}
\rho&=&\rho(\omega)\;=\;|\rho(\omega)|\,e^{{\rm i}\psi(\omega)}%
\nonumber\\%
\rho^2&=&|\rho(\omega)|^2\,e^{2{\rm i}\psi(\omega)}\;=\;%
\rho_x^2\,+\,\rho_y^2\,+\,\rho_z^2%
\label{A.17}%
\end{eqnarray}%
The spectrally dependent parameter $\rho=\rho(\omega)$ is a
combined characteristic of anisotropy effects, which can manifest
themselves in dispersion as well as in absorption.

As can be straightforwardly verified in a circular polarization
basis, the slowly varying amplitude can be expressed in terms of
phase integrals in the following form
\begin{eqnarray}
X_{q_1}{}^{q_2}({\mathbf r}_1,{\mathbf r}_2,\omega)&=&%
e^{i\phi_0({\mathbf r}_1,{\mathbf r}_2)}\,%
\left[\cos(\phi({\mathbf r}_1,{\mathbf r}_2))\,%
\delta_{q_1}{}^{q_2}\right.%
\nonumber\\%
&&\left.+\,{\rm i}\sin(\phi({\mathbf r}_1,{\mathbf r}_2))\,%
({\mathbf n}\hat{\sigma})_{q_1}{}^{q_2}\right]%
\label{A.18}%
\end{eqnarray}%
where
\begin{eqnarray}
\phi_0({\mathbf r}_1,{\mathbf r}_2)&=&\rho_0(\omega)\,\frac{2\pi\omega}{c}\,%
\int_{{\mathbf r}_2}^{{\mathbf r}_1}\chi({\mathbf r},\omega)\,ds%
\nonumber\\%
\phi({\mathbf r}_1,{\mathbf r}_2)&=&\rho(\omega)\,\frac{2\pi\omega}{c}\,%
\int_{{\mathbf r}_2}^{{\mathbf r}_1}\chi({\mathbf r},\omega)\,ds%
\label{A.19}%
\end{eqnarray}%
and
\begin{equation}
{\mathbf n}={\mathbf n}(\omega)=\frac{\mbox{\boldmath $\rho$}(\omega)}%
{\rho(\omega)}%
\label{A.20}%
\end{equation}%
The integrals in (\ref{A.19}) are evaluated along the light ray
linking the points ${\mathbf r}_2$ and ${\mathbf r}_1$. All the
parameters on the right hand side of (\ref{A.18}) are spectrally
dependent.

The Cartesian tensor components of the slowly varying amplitudes
$X_{ij}({\mathbf r}_1,{\mathbf r}_2,\omega)$ can be restored via
the transformations reverse to (\ref{A.11}) with no remaining
importance given to the covariant notation. In a Cartesian basis
set the slowly varying amplitudes are given by
\begin{eqnarray}
X_{11}({\mathbf r}_1,{\mathbf r}_2,\omega)&=&%
e^{i\phi_0({\mathbf r}_1,{\mathbf r}_2)}\,%
\left[\cos(\phi({\mathbf r}_1,{\mathbf r}_2))\right.%
\nonumber\\%
&&\phantom{e^{i\phi_0({\mathbf r}_1,{\mathbf r}_2)}\,}%
\left.-\,{\rm i}\sin(\phi({\mathbf r}_1,{\mathbf r}_2))\,%
n_x\right]%
\nonumber\\%
X_{22}({\mathbf r}_1,{\mathbf r}_2,\omega)&=&%
e^{i\phi_0({\mathbf r}_1,{\mathbf r}_2)}\,%
\left[\cos(\phi({\mathbf r}_1,{\mathbf r}_2))\right.%
\nonumber\\%
&&\phantom{e^{i\phi_0({\mathbf r}_1,{\mathbf r}_2)}\,}%
\left.+\,{\rm i}\sin(\phi({\mathbf r}_1,{\mathbf r}_2))\,%
n_x\right]%
\nonumber\\%
X_{12}({\mathbf r}_1,{\mathbf r}_2,\omega)&=&%
e^{i\phi_0({\mathbf r}_1,{\mathbf r}_2)}\,%
{\rm i}\sin(\phi({\mathbf r}_1,{\mathbf r}_2))\,(n_y+{\rm i}n_z)%
\nonumber\\%
X_{21}({\mathbf r}_1,{\mathbf r}_2,\omega)&=&%
e^{i\phi_0({\mathbf r}_1,{\mathbf r}_2)}\,%
{\rm i}\sin(\phi({\mathbf r}_1,{\mathbf r}_2))\,(n_y-{\rm i}n_z)%
\nonumber\\%
&&\label{A.21}%
\end{eqnarray}%
where the indices $1$ and $2$ relate to the $x$ and $y$ components
respectively. In these expressions, the physical meaning of the
symbolic vectors $\mbox{\boldmath $\rho$}$ or ${\mathbf n}$ is
clearly seen. The component $\rho_z$ is responsible for gyrotropy
and circular dichroism of the atomic vapor. The other two
components $\rho_x$ and $\rho_y$ are responsible for the effects
of birefringence and dichroism in linear polarization defined with
respect to either $x,y$ axes or to an alternative basis rotated
relative to $x,y$ by an angle of $\pi/4$. Let us point out that
the phase integrals (\ref{A.19}) have real (dispersion) as well as
imaginary (absorption) parts.

We conclude this appendix by the following remark concerning the
validity of the representation of slowly varying amplitudes in the
forms (\ref{A.18}) or (\ref{A.21}). As follows from the above
discussion, our approach does not apply when the spatial
distribution of any type of polarization does not directly reflect
the density distribution of atoms. In the most general case,
different polarization components can be described by different
spatial distributions. However there are two important limits
where our assumptions are self-consistent.  The first is resonant
scattering in a dense atomic cloud. Then the photon free path
scaling an average separation between the points ${\mathbf r}_1$
and ${\mathbf r}_2$, see example (\ref{2.2}), can be small enough
in comparison with the spatial scale where the distribution of the
polarization components has noticeably changed. Then in a
practical application of (\ref{A.21}), as a first and reliable
approximation, it seems reasonable to use the polarization
components averaged for the atoms located along the light ray
between these points. The second quite important limit is a highly
polarized atomic ensemble with 100\% spin orientation. For such an
ensemble the polarization distribution normalized to the local
density will be uniform with high accuracy.

\end{document}